\documentclass[sigconf]{acmart}
\usepackage{array}
\usepackage{xspace}
\usepackage{balance}
\usepackage{appendix}
\usepackage{booktabs}
\usepackage{amsmath}
\usepackage{url}
\newtheorem{definition}{Definition}
\newtheorem{theorem}{Theorem}
\newtheorem{observation}{Observation}

\newcommand{\myparagraph}[1]{\vspace{0.6\baselineskip}\noindent{\textbf{#1}}.~}
\newcommand{\firstmention}[1]{{\bf #1}}

\newcommand{\baseShort}{{\bf R}\xspace}
\newcommand{\divShort}{{\bf D}\xspace}
\newcommand{\Baseline}{Relevance\xspace}
\newcommand{\Diversity}{Diversity\xspace}

\newcommand{\omt}[1]{}
\newcommand{\numPlayers}{n}
\newcommand{\setPlayers}{N=\{1, 2, ..., n\}}

\newcommand{\definedas}{\stackrel{def}{=}}

\newcommand{\query}{q}
\newcommand{\Di}{D_i}

\newcommand{\Dj}{D_j}
\newcommand{\D}{D=\cup_{i=1}^{n}D_{i}}
\newcommand{\Dalone}{D}
\newcommand{\Vlf}[1]{V_{#1}^{(1)}}
\newcommand{\Vls}[1]{V_{#1}^{(2)}}

\newcommand{\rankingFunction}{r:D\rightarrow \mathbb{R^{+}}}
\newcommand{\rankFun}{r}
\newcommand{\RSP}{RSP(D_1, ..., D_n) = RSP(D)}
\newcommand{\PRSP}{PRSP(D_1, ..., D_n) = PRSP(D)}

\newcommand{\strategyProfile}{\dvector=(d_1,\ldots, d_n)}
\newcommand{\strategyi}{d_i}
\newcommand{\strategymi}{d_{-i}}
\newcommand{\prsp}{r_p}
\newcommand{\Regret}{REGRET_i(\strategyi, \strategymi;\prsp)}

\newcommand{\KnowledgeStatel}{R_{p}^{l}}

\newcommand{\dd}{d}
\newcommand{\ddi}{d_i}
\newcommand{\dw}{d_w}

\newcommand{\ddmi}{d_{-i}}

\newcommand{\dvector}{\underline{d}}
\newcommand{\dlvector}{\underline{d}^{l}}
\newcommand{\ddj}{d_j}

\AtBeginDocument{%
  \providecommand\BibTeX{{%
    \normalfont B\kern-0.5em{\scshape i\kern-0.25em b}\kern-0.8em\TeX}}}

\renewcommand\footnotetextcopyrightpermission[1]{} 


\makeatletter
\gdef\@copyrightpermission{
 \begin{minipage}{0.3\columnwidth}
 \href{https://creativecommons.org/licenses/by/4.0/}{\includegraphics[width=0.90\textwidth]{figs/4ACM-CC-by-88x31.eps}}
 \end{minipage}\hfill
 \begin{minipage}{0.7\columnwidth}
 \href{https://creativecommons.org/licenses/by/4.0/}{This work is licensed under a Creative Commons 
Attribution International 4.0 License.}
 \end{minipage}
 \vspace{5pt}
}
\makeatother

\begin{document}

\title{Search results diversification in competitive search}

\author{Tommy Mordo}
\affiliation{
 \institution{Technion}
 \city{Haifa}
 \country{Israel}
}
\email{tommymordo@technion.ac.il}

\author{Itamar Reinman}
\affiliation{
 \institution{Technion}
 \city{Haifa}
 \country{Israel}
}
\email{itamarr@campus.technion.ac.il}

\author{Moshe Tennenholtz}
\affiliation{
 \institution{Technion}
 \city{Haifa}
 \country{Israel}
}
\email{moshet@technion.ac.il}

\author{Oren Kurland}
\affiliation{
 \institution{Technion}
 \city{Haifa}
 \country{Israel}
}
\email{kurland@technion.ac.il}

\renewcommand{\shortauthors}{Tommy Mordo, Itamar Reinman, Moshe Tennenholtz \& Oren Kurland}

\begin{abstract}
In Web retrieval, there are many cases of competition between authors of Web documents: their incentive is to have their documents highly ranked for queries of interest. As such, the Web is a prominent example of a competitive search setting. Past work on competitive search focused on ranking functions based solely on relevance estimation. We study ranking functions that integrate a results-diversification aspect. We show that the competitive search setting with diversity-based ranking has an equilibrium. Furthermore, we theoretically and empirically show that the phenomenon of authors mimicking content in documents highly ranked in the past, which was demonstrated in previous work, is mitigated when search results diversification is applied.
\end{abstract}

\maketitle
\section{Introduction}
Competitive search \cite{kurland_competitive_2022} is a retrieval
setting where some document authors, henceforth referred to as publishers, are
ranking incentivized. That is, their goal is to have their documents
highly ranked for queries of interest. For example, in Web search, the
highest ranks on the first page of results attract most clicks
\cite{Joachims+al:05a}. Hence, there is often a ranking competition for queries of commercial intent.
Competition for high ranks can bring about unwarranted publishers'
actions which hurt users, and consequently the search ecosystem;
e.g., spamming \cite{Gyongyi+Molina:05a,Castillo+Davison:10a}. These
actions are often referred to as black hat search engine optimization
(SEO) \cite{Gyongyi+Molina:05a}. In contrast, white hat search engine
optimization, which is the focus of previous work on competitive
search \cite{kurland_competitive_2022}, as well as ours here, refers
to legitimate actions; specifically, document modifications intended to improve document ranking and which do not hurt document quality and/or the search ecosystem\footnote{Competitive search can be viewed as part or aspect of a field often referred to as adversarial retrieval \cite{Castillo+Davison:10a}.}. 

The competitive search setting can be modeled as a game
\cite{kurland_competitive_2022}: publishers are players; their actions
are document modifications applied in response to induced ranking to improve future ranking. Hence, it is only natural to use
game theory to model the setting; specifically, so as to answer
questions about whether the competition reaches a steady state
(equilibrium) and what are prevalent publishers' strategies.  For example, Ben Basat et al. \cite{Basat+al:17a}
showed that if the ranking function is disclosed to publishers, then
the probability ranking principle \cite{Robertson:77a} --- which is
the underlying pinning of most relevance ranking functions --- is
sub optimal. Specifically, there are stochastic ranking functions that lead to broader topical diversity in the long run in
the corpus \cite{Basat+al:17a}. This theoretical finding leads to an highly important
observation: while the standard practice of evaluating search systems
is measuring performance of the ranked list, there are long-term
corpus effects driven by ranking incentives which should also be
analyzed and accounted for.

In reality, ranking functions are not disclosed to
publishers. Raifer et al. \cite{raifer_information_2017} then analyzed the
competitive search setting as a {\em repeated game}
\cite{aumann1995repeated}: publishers continuously respond to
rankings induced by a function they do not know to improve
future ranking. They found that these repeated games reach a so called
{\em min-max regret equilibrium} which is a stable state \cite{Hyafil+Boutilier:12a}. Furthermore, Raifer et al.'s \cite{raifer_information_2017}
analysis revealed a ``mimicking the winner'' document modification
strategy:  publishers modify their documents by mimicking content in the documents most highly ranked (``winners'') in the past for the same query. 
Since the ranking function is undisclosed, highly ranked documents are a ``signal'' about what the ranking function rewards. Raifer et al. \cite{raifer_information_2017} also organized ranking competitions between students where ``mimicking the winner'' clearly emerged as a prevalent strategy.

Goren et al. \cite{goren_driving_2021} empirically showed that the
``mimicking the winner'' strategy in
ranking competitions results in a {\em herding} phenomenon which
was studied in the economics literature
\cite{Banerjee,Bikhchandani,SmithSorensen}. Specifically, they
organized ranking competitions where they planted documents at the
highest ranks. These documents manifested various effects. For
example, they were non-relevant to the query or emphasized one query
aspect but did not touch on another query aspect. The players in Goren
et al.'s ranking competitions \cite{goren_driving_2021} applied the
``mimicking the winner'' strategy as was the case in Rafier et al.'s
competitions \cite{raifer_information_2017}. The resultant documents
manifested the same effects as those manifested by the planted documents: non-relevance and emphasis on one query
aspect without any coverage of another. In other words, publisher herds were
formed. Goren et al. \cite{goren_driving_2021} warned that aside for
reducing diversity in the corpus, the herding effect could be exploited by
publishers interested in driving various unwarranted phenomena in the corpus. 

The findings about herding in ranking competitions motivate our main
research question in this paper: {\em how can we reduce the extent to
 which the ``mimicking the winner'' strategy is applied in
 competitive search so as to ameliorate the herding effect?''} A
supposedly obvious approach is to apply search results diversification
\cite{Santos+al:15a,wu_result_2024}. For example, in the classical maximal marginal
relevance (MMR) retrieval method \cite{carbonell_use_1999}, the
retrieval score of a document is penalized to the extent the document
is similar to documents already ranked higher. Applying
diversity-based ranking in competitive search --- i.e., using both
relevance estimates and search-results diversification --- gives rise to the following research questions which we address in this paper: (i)
does diversity-based ranking result in an equilibrium? (ii) what are
the players' document modification strategies? and, (iii) is ``mimicking the winner'' strategy ameliorated?

To address these questions, we present the first (to the best
of our knowledge) theoretical analysis of the competitive search
setting with diversity-based ranking. We analyze the resultant
repeated ranking game and prove that there is a min-max regret
equilibrium (question (i)). In doing so, we show that some players
(publishers), as from a certain point, will cease to compete for the
first rank position and will focus on trying to secure the second rank
position (question (ii)). To do so, their documents should not be very similar to the
highest ranked ones due to the diversity-based ranking. As a result,
the ``mimicking the winner'' strategy is less prevalent helping to
ameliorate the extent of herding (question (iii)).

We provide empirical support to our game theoretical
findings. Specifically, we organized ranking competitions where we
used both (i) rankings based solely on relevance estimation as was the
case in past ranking competitions
and (ii) rankings based on both relevance estimation and
search-results diversification. Our ranking competitions are also the
first, to the best of our knowledge, to apply dense retrieval with
document and query embeddings. Analysis of the competitions revealed
that when diversity-based ranking was used, the ``mimicking the
winner'' strategy was applied to a  reduced extent than
when ranking was based solely on relevance estimation. We also found
that diversity-based ranking resulted in increased content diversity with respect to ranking based solely on relevance estimation. Together, these findings attest that diversity-based ranking helps to ameliorate herding of publishers.

The dataset of the competitions, and the accompanying code, will be made public upon publication of this paper. They are available for reviewing purposes at \url{https://github.com/diversityamelioratingherding/dataset} and \url{https://github.com/diversityamelioratingherding/code}.

Our contributions can be summarized as follows:
\begin{itemize}
\item The first theoretical analysis, to the best of our knowledge, of a competitive search setting where diversity-based ranking is applied.
  \item Two important theoretical results for ranking games with
    diversity-based ranking: (i) they have a min-max regret
    equilibrium, and (ii) some players (publishers) focus on securing the second rank position and hence do not mimic the highest ranked document (winner). Hence, the herding effect observed in past work is ameliorated.
  \item Organizing the first ranking competitions with diversity-based ranking. The analysis of these competitions provides support to our theoretical result about a reduced application of the ``mimicking the winner'' strategy. Together with findings about content diversity, we provide empirical support to the fact that diversity-based ranking helps to ameliorate publisher herding.
    \item A public dataset (with accompanying code) of the ranking competitions we organized.
  \end{itemize}


\section{Related Work}
\label{sec:rel}
There is a large body of work on methods for diversifying search
results \cite{Santos+al:15a,wu_result_2024}. Our goal is to analyze
how diversification affects the competitive search setting.

Prior work on using game theory to analyze the competitive retrieval
setting has focused on ranking functions that only use relevance
estimates without search-results diversification
\cite{Basat+al:17a,raifer_information_2017,nachimovsky_ranking-incentivized_2024}. Ben Basat
et al. \cite{Basat+al:17a} assumed knowledge of the ranking function,
and hence the resultant games were of complete and perfect
information; accordingly, they analyzed the Nash equilibrium. Raifer et
al. \cite{raifer_information_2017} assumed an undisclosed ranking
function and analyzed the min-max regret equilibrium of a repeated
game as we do here. In contrast to our work, they used a ranking
function which does not apply diversification. Nachimovsky et
al. \cite{nachimovsky_ranking-incentivized_2024} characterized the cases where a Nash equilibrium exists in a competitive setting where a publisher modifies a
document for several queries representing an information
need. Extending our analysis of diversity-based ranking to a setting where publishers compete for multiple queries is left for future work.

There is work on recommendation systems \cite{eilat_performative_2023}
that shows how learning algorithms can incentivize strategic content
creators (publishers) to produce diverse content. In contrast to our
work, no (game) theoretic analysis was reported.

There is a recent line of work on devising specific (algorithmic)
adversarial attacks to promote documents in rankings
\cite{goren_ranking-incentivized_2020,Jenkins+al:20a,Raval+Verma:20a,Song+al:20a,Song+al:22a,Wang+al:22a,Wu+al:22a,Liu+al:23a,Xuanang+al:23a}. In
contrast, we address a setting where humans modify documents so as to
improve their ranking with no information about the underlying ranking
function. There is also work on improving the robustness of retrieval methods to adversarial attacks \cite{Wu+al:22b,Vasilisky+al:23a,Liu+al:24a}.



\section{Game Theoretic Analysis}
\label{sec:model}
The ranking competition between publishers (document authors) is
driven by their ranking incentives \cite{kurland_competitive_2022}. That is, we assume that some
publishers opt to have their documents highly ranked for a given
query. In response to a ranking induced for the query, the publishers
might modify their documents so as to improve their future ranking. In
what follows we analyze this on-going ranking game as a {\em repeated game} \cite{aumann1995repeated}. Previous work on using game theory to analyze ranking competitions assumed that the sole criterion for ranking is relevance estimation applied independently for documents \cite{Basat+al:17a,raifer_information_2017,nachimovsky_ranking-incentivized_2024}. We analyze ranking games where the ranking function employs also search-results diversification \cite{Santos+al:15a}.



\subsection{The ranking game}
We assume a fixed query $\query$ and a fixed ranking function defined
below. Let $\setPlayers$ be a set of $\numPlayers$ publishers who are
the players in a repeated ranking game. Each player is incentivized to
attain a high ranking for her document in response to $\query$ every
round\footnote{A round corresponds to the event of publishing documents and ranking them
 in response to a query.}. We define the {\em strategy profile} of a
round in a repeated game, $\dvector$, as the set of documents
published by all the players in that round; $\dlvector$ stands for the
strategy profile in round $l$. We sometimes write $\dvector$ as
($\ddi, \ddmi$) to emphasize the document published by player $i$
($\ddi$) and the set of all other players' documents ($\ddmi$).

Let $\Di$ be a finite and fixed set of documents that player $i$ may
produce in any round of the game; each player produces a single
document per round. We assume $\Di \cap \Dj = \varnothing$ for every
$i, j$. The collection of all documents that can be produced by the
players is $\D$. In every round of the game, the players'
documents are ranked with respect to the query; observing the ranking, the players may modify
their documents to improve their future rankings.

As in previous work on analyzing ranking games \cite{raifer_information_2017}, we assume a complete linear ordering over $D$, denoted ``<''. The
ordering can be based on a single numeric feature in the document
representation or the similarity to a reference document or a query.
To facilitate
the exposition, we associate $\Dalone$ with elements in $[0,1]$.

A retrieval (ranking) function is a mapping $\rankingFunction$
that assigns a non-negative retrieval score to a document; usually, the score is a relevance estimate or a proxy thereof. To simplify the mathematical analysis, we 
assume no retrieval-score ties: $r(d_i) \neq r(d_j)$ for any
$d_i, d_j \in D$ where $d_i \neq d_j$. Inspired by Raifer et al.'s \cite{raifer_information_2017} analysis of repeated ranking games, we assume a single peak ranking function:
\begin{definition}
Let $\RSP$ denote the set of all possible single peak ranking functions. These functions satisfy the condition that for any $d \in D$, there do not exist $\ddi, \ddj \in D$ for which $\ddi < \dd < \ddj$ and $r(\ddi) > r(d)$ and $r(\ddj) > r(d)$ hold.
\end{definition}

Although many effective ranking functions, such as non-linear
feature-based learning-to-rank \cite{Liu:11a} or neural methods
\cite{mitra_introduction_2018}, are not single peak, we emphasize that
our analysis adopts the potential perspective of documents'
publishers (players) to whom the ranking function is undisclosed. Specifically, document modification (e.g., adding query terms
to the document) can help to improve ranking up until a point
where the modifications lead to retrieval score penalty as they are considered excessive
and/or harming document quality. A case in point, feature-based
learning-to-rank methods applied over the Web often include
query-independent document quality estimates \cite{Bendersky+al:11a} (e.g., spam estimates) alongside features that quantify surface-level document-query similarities (e.g., BM25 score). Increasing query-terms occurrence in the document increases this surface-level similarity and hence increases retrieval score; however, adding more query terms can, as from a certain point, decrease the retrieval score due to the document quality estimates (e.g., having the spam score increase).


In previous work on analyzing ranking games
\cite{raifer_information_2017, nachimovsky_ranking-incentivized_2024},
a document was assigned a retrieval score independently of other
documents (cf., the probability ranking principle
\cite{Robertson:77a}); the retrieval score was assumed to rely on a
relevance estimate. Our goal is to analyze ranking functions that
apply {\em results diversification} \cite{Santos+al:15a} where
documents' retrieval scores can be dependent on each
other. Specifically, we assume an iterative retrieval-score assignment
procedure as in Maximal Marginal Relevance (MMR)
\cite{carbonell_use_1999}, where the retrieval score of a document
depends also on the document similarity to documents already ranked
above it. Specifically, in MMR \cite{carbonell_use_1999}, the retrieval
score of a document $d$ which was not positioned yet in the ranked list for query $q$ is
\begin{equation}
  \label{eq:mmr}
  Score_{MMR}(d) \definedas (1-\lambda) s(d,q) - \lambda \max_{d' \in T} sim(d,d'),
\end{equation}
  where $s(d,q)$ is a basic relevance estimate, $T$ is a set of documents already positioned at the highest ranks, $sim(\cdot,\cdot)$ is an inter-document similarity measure and $\lambda$ is a free parameter.

Previous work on analyzing repeated ranking games using game theory
focused on the highest ranked document
\cite{raifer_information_2017}. Since we explore diversity-based
ranking, our game theoretic analysis focuses WLOG on the top-two
ranked documents: the second document is selected based
also on its similarity with the highest ranked document.

We
therefore define the retrieval score assigned to documents in the
corpus only after the highest ranked document was selected using the
basic retrieval function $\rankFun$. As in MMR \cite{carbonell_use_1999}, we penalize the retrieval score
of documents which are highly similar to the selected document:
\begin{definition}\label{definition_prsp}
Let $\PRSP$ denote the set of all ranking functions $\prsp$ with the
following property. The highest ranked document  $\dd^{*} \definedas \arg max_{d' \in
\{d_i\}_{i=1}^{n}} r(d')$ is selected using some basic single peak ranking function $r$.   Then, for any document $d_i$ where $ |\dd^{*} - d_i| \ge \alpha$, $d_i$'s
retrieval score is $\prsp(\dd_i) \definedas r(d_i)$.
For any document $d_i$ where $0 < |\dd^{*} - d_i| < \alpha$, $d_i$'s
retrieval score, $\prsp(\dd_i)$, is lower than $r(d_j)$ for all 
documents $d_j$ for which $|\dd^{*} - d_i| \ge \alpha$; $\alpha$ is a free parameter.

%


\end{definition}
In other words, functions in $PRSP(D)$ penalize the retrieval scores
of documents whose similarity to the highest ranked document, $d^{*}$, is above
$\alpha$ to an extent that these documents are then ranked lower than all documents whose similarity to $d^{*}$ is below $\alpha$. MMR \cite{carbonell_use_1999}, with a relatively high value of $\lambda$ in Equation \ref{eq:mmr}, can serve as an example of such function. Another simple example of such a function is that which assigns document $d$ a negative score of $r(d)-r(\dd^{*})$ in case $|\dd^{*} - d| < \alpha$ and $r(d)$ otherwise.
Note that since the basic ranking function $r$ is single peak, so are the functions in $PRSP(D)$ as the selection of the highest ranked document depends solely on $r$.
We also assume WLOG that as is the case for $r$, functions in $PRSP(D)$ yield no retrieval-scores ties.

Since we focus on rankings and not retrieval scores used to induce
them, herein we refer to the set of ranking functions $PRSP(D)$ from
Definition \ref{definition_prsp} also as the set of all possible
rankings (i.e., total orderings) induced over $D$; i.e., those induced
by the functions in $PRSP(D)$. Players gain knowledge throughout the
game by observing rankings induced over documents. Specifically, they can infer that some ranking functions (ordering) are not possible. We use $R^{l}_{p} \subseteq PRSP(D)$ to
denote the possible subset of $PRSP(D)$ as inferred by the players at the beginning of round $l$; we refer to $R^{l}_{p} \subseteq PRSP(D)$ as the {\em knowledge state} in round $l$. We
assume players in the ranking game are rational: (i) they are motivated to have their documents ranked
as high as possible, (ii) they continuously learn the knowledge state in
each round; i.e., the set of ranking functions that could have induced the rankings they observed. Consequently, the knowledge state can only shrink in size
over time: $R^{l+1}_{p} \subseteq R^{l}_{p}$.

Previous work on analyzing ranking games focused solely on the highest
ranked document
\cite{raifer_information_2017,nachimovsky_ranking-incentivized_2024},
specifically, in defining player utility; i.e., players not ranked
first received zero utility. Since we address diversification, we
attribute WLOG non-zero utility to the publishers of the two highest ranked documents:


\begin{definition}
The utility of player $i$ in round $l$ assuming a ranking function $\prsp$ (Definition \ref{definition_prsp}) is defined as:
$$U_{i}^{l}(\dd_i, \ddmi;\prsp) =  
\begin{cases}
    1 & \dd_i \text{ is ranked first};\\ \beta & \dd_i \text{ is ranked second}; \\ 0 &\text{otherwise};
\end{cases}$$
\end{definition}
$\beta < 1$ is a free parameter.
As in Raifer et al. \cite{raifer_information_2017}, the utility of a player over $t$ rounds is the sum of her per-round utility given the strategy profile at each round: 
\begin{definition}
$U_{i}(\{\dlvector\}_{l=1}^{t};\prsp) \definedas \sum_{l=1}^{t} U_{i}^{l}(d_i^l,d_{-i}^{l};\prsp)$; $\dlvector = (d_i^l,d_{-i}^{l})$ is the strategy profile of round $l$ where $d_i^l$ and $d_{-i}^{l}$ are the documents published by player $i$ in round $l$, and by all other players in round $l$, respectively.
\end{definition}

To account for the effort required to modify a document, we introduce
a negligible cost $C$ associated with a document modification. The
utility in a given round will be adjusted by subtracting the cost of
changes, assuming $C|D| < \beta$ and $eC > 0$ for a change of
distance $e$\footnote{Recall that documents correspond to elements in
  $[0,1]$. Hence, document similarity is measured by the difference
  between the respective elements.}. 

We define the set of possible documents, determined by the knowledge state of round $l$, that might be ranked first or second:

\begin{definition}\label{definition_vlf}
Let $\Vlf{l} \subseteq [0,1]$ be the set of documents that can be
ranked first by the functions in the knowledge state $R^{l}_{p}$ at
round $l$; i.e., for these documents the functions attain the peaks.
\end{definition}

\begin{definition}\label{definition_vls}
Let $\Vls{l} \subseteq [0,1]$ be the set of documents that can
be ranked second by the ranking functions in the
knowledge state $R^{l}_{p}$, assuming that the
highest ranked documents are selected from $\Vlf{l}$ in Definition \ref{definition_vlf}.
\end{definition}

The set $\Vlf{l}$ reflects the uncertainty about the peaks of the
ranking functions in $PRSP(D)$. The set $\Vls{l}$ reflects uncertainty
not only about the peaks of functions in $PRSP(D)$, but also about the
similarity threshold $\alpha$ used for diversity-based score
penalty. (See Definition \ref{definition_prsp}.) Note that $|D \cap \Vlf{l}| \leq |D \cap \Vls{l}|$: there are more documents that can be ranked second than those which can be ranked first since the ranking functions are single peak.


A central challenge in analyzing repeated games
\cite{aumann1995repeated} is identifying a suitable {\em solution
  concept} that characterizes the strategic behavior of players. The
Nash equilibrium for example, which is a fundamental solution concept
in game theory where no player benefits from unilaterally deviating
from her strategy, is unsuitable in our setting (cf., \cite{raifer_information_2017}). Specifically, the repeated
game we address has incomplete and imperfect information: the ranking function is undisclosed (incomplete information) and players do not know the documents published by other players for the next ranking to be induced (imperfect information). Consequently, we use the {\em minmax regret equilibrium} \cite{Hyafil+Boutilier:12a} as an alternative solution concept.

\subsection{Minmax regret equilibrium}
In minmax regret equilibrium, each player simultaneously selects a
strategy (a document in our setting) that minimizes her regret with respect to her best
response\footnote{Best response is the strategy with the highest utility a player can play
given her assumption on the strategies of all other players.} assuming she had knowledge of the ranking function and that
all other players stick to their strategies. We begin by formally
defining {\em regret}:

\begin{definition}
Given a strategy profile $\strategyProfile$ and a ranking function $\prsp$ from Definition \ref{definition_prsp}, the regret of player $i$ from publishing document $d_i$ is:
$$\Regret \definedas \max_{x \in D_i}U_i(x, \strategymi; r_p) -
  U_i(\strategyi, \strategymi; r_p).$$ Note that $argmax_{x \in
    D_i}U_i(x, \strategymi; r_p)$ is $i'$s best response to the
strategies of all other players, $\strategymi$.

\end{definition}
This regret is the maximum gain a player can attain by deviating from
the given strategy ($d_i$) assuming she knows the ranking function
$\prsp$. The maximal regret over all possible ranking functions $\prsp$ ($\in PRSP(D)$) is:
\begin{definition}
The maximal regret with respect to $\strategyi$ is:
  $$MR_i(\strategyi, \strategymi; PRSP(D)) \definedas max_{r_p \in PRSP(D)}\Regret.$$

\end{definition}
A minmax regret equilibrium is defined as:
\begin{definition}
A strategy profile $\strategyProfile$ is a minmax regret equilibrium if for every player $i$:\\ $d_i = argmin_{x \in D_i}MR_i(x, \strategymi; PRSP(D))$.
\end{definition}

We now arrive to a fundamental result about the stability (i.e., equilibrium) of repeated ranking games in our setting:
\begin{theorem} \label{theorem_minmax}
A repeated ranking game in our setting has a minmax regret equilibrium in every round.
\end{theorem}
\begin{proof}
We construct the minmax regret equilibrium in the game for round $l$.
Let $\KnowledgeStatel$ be the knowledge state at the beginning of round $l$. In round $l=0$ all ranking functions $\prsp$ in $PRSP(D)$ are possible.  In each of the following rounds, the knowledge state size can only shrink. Let $d_{i}^{l-1}$ be the document selected (published) by player $i$ in round $l-1$. We define the document $d_{i}^{l}$ to be published in round $l$ using Definitions \ref{definition_vlf} and \ref{definition_vls} of $\Vlf{l}$ and $\Vls{l}$, respectively: the documents that can be ranked first and those that can consequently be ranked second at the beginning of round $l$.
\begin{itemize}
    \item If $\Di\cap \Vlf{l} \neq \varnothing$ then we select $d_{i}^{l} \in \Di\cap \Vlf{l}$ s.t. $|d_{i}^{l} - d_{i}^{l-1}|$ is minimal. (Ties are broken arbitrarily.)
    \item If $\Di\cap \Vlf{l} = \varnothing$ then:
    \begin{itemize}
        \item If $\Di\cap \Vls{l} \neq \varnothing$ then we select $d_{i}^{l} \in \Di\cap \Vls{l}$ s.t. $|d_{i}^{l} - d_{i}^{l-1}|$ is minimal. (Ties are broken arbitrarily.)
        \item If $\Di\cap \Vls{l} = \varnothing$ then we define $d_{i}^{l} = d_{i}^{l-1}$.
    \end{itemize}
\end{itemize}
We now show that the strategy profile ($d_{1}^{l}, \ldots, d_{n}^{l}$) is a minmax regret equilibrium of the game in round $l$.

Consider player $i$. No player $j \neq i$ will publish a document not
in $\Vlf{l} \cup \Vls{l}$, as doing so would prevent $j$ from being
ranked first or second. Thus, for player $i$, publishing a document
not in $\Vlf{l} \cup \Vls{l}$ is dominated by simply re-publishing
their previous document. Since any document in $\Vlf{l}$ or $\Vls{l}$
can potentially secure the first or second rank position, the highest
regret for player $i$ is for not publishing a document
from $\Vlf{l}$ or $\Vls{l}$. In terms of regret, selecting a
document from $\Vlf{l}$ is preferable to selecting from $\Vls{l}$, since the utility for the first rank position is higher than that of
the second. By induction, this logic applies to every player
$i$ and round $l$. Since player $i$ can only publish documents in
$D_i$, she will strive to publish a document from $D_i \cap \Vlf{l}$ or
$D_i \cap \Vls{l}$ with minimal modification cost.
\end{proof}

The construction of minmax regret equilibrium just presented gives rise to the following corollary and observation. Herein we refer to the highest ranked document in a round as a {\em winner}, denoted $d_w$.

\begin{corollary}
  \label{cor:strategy}
  Players who did not win in round $l-1$ will publish a document in round $l$ that tends to become more similar to either (i) the winning document $\dw$ of round $l-1$, \textbf{or} (ii) what they assume to be the second highest ranked document, while minimizing the cost of modifying their previous document. If they cannot attain the first or second rank position, they will republish their previous document.
\end{corollary}

\begin{proof}
Assume, without loss of generality, that a player $i$ who did not win
in round $l-1$ published a document $d_{i}^{l-1}$ that satisfies
$d_{i}^{l-1} \leq \min_{d_j\in \Vlf{l}} d_j \leq \dw \leq \max_{d_j\in \Vlf{l}} d_j$
    where $\dw$ is the winning document of round $l-1$ (i.e., ranked the highest). We consider four cases:

\begin{itemize}
\item If the interval $D_i \cap [\min_{d_j\in \Vlf{l}} d_j, \dw]$ is not empty, by Theorem \ref{theorem_minmax} player $i$ will publish a document $d \in D_i \cap [\min_{d_j\in \Vlf{l}} d_j, \dw]$ that minimizes $|d_{i}^{l-1} - d|$. Player $i$ will avoid publishing any document $d \in D_i \cap [\dw,\max_{d_j\in \Vlf{l}} d_j ]$ because the regret from publishing in this interval would be higher, as the cost of changing the document is larger. Thus, $d_{i}^{l-1} \leq d \leq \dw$ which means that the next document player $i$ will publish is more similar to the winner $\dw$ of round $l-1$ than her current document.

    \item If the interval $D_i \cap [\min_{d_j\in \Vlf{l}} d_j , \dw]$ is empty, by Theorem \ref{theorem_minmax} player $i$ will publish a document $d \in D_i \cap [\dw, \max _{d_j\in \Vlf{l}} d_j]$ that minimizes $|d_{i}^{l-1} - d|$. In this case, $d$ will be the most similar document to $\dw$ in $D_k$.

    \item If the interval $D_i \cap \Vlf{l}$ is empty, then by Theorem \ref{theorem_minmax}, in round $l$ player $i$ will select a document $d \in D_i \cap \Vls{l}$ that minimizes $|d_{i}^{l-1} - d|$.

    \item If both $D_i \cap \Vlf{l}$ and $D_i \cap \Vls{l}$ are empty, player $i$ will publish the same document she published in the previous round, $d_{i}^{l-1}$, to minimize modification cost.
\end{itemize}
\end{proof}

Corollary \ref{cor:strategy} leads to the following observation:
\begin{observation} \label{observation_Ti}
  For every player $i$ whose document set $\Di$ does not include the
  one for which the ranking function has a peak, there is a round as
  from which $i$ will aim for the second rank position or will simply
  continue to publish the same document.

\end{observation}





The corollary and the observation help to elucidate a key strategy of
the players: in diversity-based ranking functions, we expect to see a
mitigation of the "mimicking the winner" phenomenon \cite{raifer_information_2017} as the number of
rounds increases. That is, fewer documents become highly similar to those highly ranked in the past. The reason is that at some point of the
game, players instead of mimicking the top-ranked document, aim to secure the
second-best position. When the diversity aspect (e.g., penalty as in Definition \ref{definition_prsp}) is non-negligible, the set of
documents expected to rank second differs significantly from those
ranked first.

\section{Empirical Analysis}
\label{sec:eval}
Our next order of business is studying empirically the strategic
behavior of players in a repeated ranking game where the ranking
function applies diversification. In Section \ref{sec:data} we describe the dataset we analyzed which is a result of running ranking competitions. In Section \ref{sec:analysis} we present analysis of the dataset.

\subsection{Dataset}
\label{sec:data}
There are a few datasets which are the result of running ranking
competitions
\cite{raifer_information_2017,goren_ranking-incentivized_2020,goren_driving_2021,nachimovsky_ranking-incentivized_2024}. However,
the ranking in these competitions was solely based on
relevance estimation without accounting for diversification. Hence, we
organized similar ranking competitions that include diversity-based ranking.

Specifically, we organized two types of repeated ranking competitions. The first, denoted {\em \Baseline} (\baseShort), was based on using a dense retrieval approach
with no results diversification \cite{wang_text_2024}. In
the second type of competition, denoted {\em \Diversity} (\divShort), a diversification method was applied in
addition to the dense retrieval approach.


Forty students in an information retrieval course were the players in
the competitions. They served as publishers and modified
documents to have them highly ranked for queries. The two types
of competitions (\baseShort and \divShort) were held separately for each of
15 queries from the TREC9-TREC12 Web tracks\footnote{The
  queries were: 9, 17, 29, 34, 45, 48, 59, 69, 78, 98, 167, 180, 182,
  193, 195.}. These queries were selected as they had commercial
intent; hence, they were likely to steer a dynamic competition;
cf., \cite{raifer_information_2017}. Each player was assigned to three
randomly selected repeated games (i.e., three different queries), where at
least one was of type \textbf{R} and one was of type \textbf{D}. No pair of students was assigned to the same competition (i.e., query and competition type) more than
once. The players were not informed that competitions were of two types (i.e., different rankers).



Each competition for a query lasted for 7 rounds. Four students
competed in each round for a query for each of the type \baseShort and
type \divShort competitions.
Before the first round, for each query, students were provided with the same initial document relevant to the query. The initial documents were created as
follows. For five queries, they were selected from a previous
ranking competition \cite{goren_driving_2021}. To generate initial
documents for the other ten queries, we used the procedure applied
in Raifer et al. \cite{raifer_information_2017}. First, we used the
query in a commercial search engine and selected a highly ranked
page. We then extracted from this page a candidate paragraph of length up to 150
words. Three annotators then judged the relevance of the passages. We
repeated the extraction process for each query until a paragraph was
judged relevant by at least two annotators. The selected paragraph was
then used as the initial document for the query for all
students.

As from the second round, students were shown the induced ranking including documents' content. They could modify their documents so as to improve their
next round ranking. Documents were plaintext and of at most $150$
words. The students were instructed to produce high quality documents
(e.g., to avoid using excessive keyword stuffing) which were relevant
to the queries; they were also asked not to use GenAI tools to modify
the documents. To incentivize students to compete for higher documents
rankings, we assigned course bonus points based on their performance
in each round\footnote{The bonus was ${0.7}$, ${0.7}/{2}$, ${0.7}/{3}$ and $0$ points for the highest, second highst, third and fourth ranked documents, respectively.}. Two ethics committees approved the
competitions (international and institutional). Each student who participated in the competition signed a
consent form and had the option to opt out at
any time. Furthermore, the students could receive a perfect grade in
the course without participating in the competitions.

\myparagraph{Ranking functions} The ranking function in the \Baseline
(\baseShort) competitions was the cosine between the (unsupervised) E5
embedding \cite{wang_text_2024}\footnote{We used the
  intfloat/e5-large-unsupervised version from the Hugging Face
  repository
  (\url{https://huggingface.co/intfloat/e5-large-unsupervised}).} of a
  query and a document. In the \Diversity (\divShort) competitions, we
  used the MMR \cite{carbonell_use_1999} ranking function from
  Equation \ref{eq:mmr}; $s(d,q)$, the basic retrieval score, was the
  min-max normalized (across documents in a round) retrieval score
  used in the \baseShort competitions (i.e., E5-based); the similarity
  between two documents, $sim(\cdot,\cdot)$, was the min-max normalized
  (across all pairs of documents in a round) cosine between their
  E5-based sentence embeddings\footnote{We used Sentence-Transformers
    from the Hugging Face
    repository (\url{https://huggingface.co/sentence-transformers})
    with the E5 encoder.}. We set $\lambda=0.5$ in MMR to have equal
  importance for relevance and diversity. Retrieval scores ties were
  arbitrarily broken.

\myparagraph{Relevance and quality judgments} Each document was judged
for binary relevance to a query by five crowd workers on the Connect
platform via CloudResearch
\cite{noauthor_introducing_2024}. Additionally, five workers annotated
the content quality of each document with the labels: valid, keyword
stuffed, or spam
\cite{raifer_information_2017,nachimovsky_ranking-incentivized_2024}. All
workers were native English speakers.

The inter-annotator agreement rate, measured using free-marginal
multi-rater Kappa, for relevance judgments in the \baseShort and
\divShort competitions was $62.3$\% and $61.7$\%, respectively. For
quality judgments, the agreement rate was $32$\% for \baseShort and
$39$\% for \divShort.

About $86\%$ and $75\%$ of the documents were
marked relevant by at least three or four out of five annotators,
respectively; this proportion is consistent with earlier findings in
single-query and multi-query ranking competitions
\cite{goren_driving_2021, nachimovsky_ranking-incentivized_2024,
  raifer_information_2017}.
  
For each document, the final quality grade was defined as the number
of annotators who judged the document as valid. Similarly, the final
relevance grade was the number of annotators who
marked the document as relevant. The NDCG@4 (NDCG of the top-4
documents)\footnote{Recall that for each query and round there were four
  competing documents.} across rounds was between $0.92$ and $0.96$
for both the \baseShort and \divShort competitions. There were no
statistically significant differences\footnote{Herein, statistical
  significance is measured using a two tailed paired permutation test with $p = 0.05$ and
  $10,000$ permutations.} between the NDCG@4 for \baseShort and
\divShort. In both competitions relevance estimates are used; in the \divShort competitions diversity is also applied. These high values of NDCG@4 are consistent with
those reported for past ranking competitions \cite{goren_driving_2021,
  nachimovsky_ranking-incentivized_2024, raifer_information_2017}. We
note that the primary focus of our study is on
ranking-incentivized manipulation strategies rather than on ranking
effectiveness.

About $77\%$ of the documents were judged valid by at least three
annotators, and about $57\%$ were judged valid by at least four annotators. These numbers are a bit lower than those reported in previous studies
\cite{goren_driving_2021, nachimovsky_ranking-incentivized_2024,
  raifer_information_2017}.



The resulting dataset of the competitions 
includes: (1) 840 documents (497 unique documents); (2) relevance judgments; and, (3) quality
annotations. The dataset is publicly
available at \url{https://github.com/diversityamelioratingherding/dataset}.

\subsection{Competition analysis}
\label{sec:analysis}
We next present analysis of the competition dataset.

\begin{table}[t]
  \setlength{\tabcolsep}{3pt}
  \caption{\label{tab:trans} The percentage of cases (queries and rounds) for rank $i$ where a document ranked $i$ at round $t$ 
    was ranked $j$ in round $t+1$ in the \Baseline and
    \Diversity competitions. The highest rank is $1$.}
  \begin{tabular}{ccccc} \\ \toprule
  \multicolumn{5}{c}{\bf \Baseline} \\
      rank@t$\backslash$ rank@(t+1) & 1 & 2 & 3 & 4 \\
      \midrule
       1 & $68\%$ & $26\%$ & $3\%$ & $3\%$ \\
           2 & $7\%$ & $46\%$ & $42\%$ & $6\%$ \\
          3 & $17\%$ & $20\%$ & $34\%$ & $29\%$ \\
          4 & $9\%$ & $9\%$ & $20\%$ & $62\%$ \\
          \bottomrule
    \multicolumn{5}{c}{\bf \Diversity} \\ 
          rank@t$\backslash$ rank@(t+1) & 1 & 2 & 3 & 4 \\ \midrule
    1 & $63\%$ & $11\%$ & $23\%$ & $2\%$ \\
    2 & $11\%$ & $58\%$ & $13\%$ & $18\%$ \\
    3 & $17\%$ & $12\%$ & $40\%$ & $31\%$ \\
    4 & $9\%$ & $19\%$ & $23\%$ & $49\%$ \\
   \end{tabular}


  \end{table}

\subsubsection{Changes of rank positions}
Our first order of business is analyzing the rank changes of documents in
the two types of competitions: \Baseline (\baseShort) and \Diversity (\divShort).
In Table \ref{tab:trans} we report for each rank $i$ ($\in \{1,\ldots,4\}$) the percentage of
cases (with respect to queries and rounds) where a document ranked $i$ in some round $t$ ($\in \{1,\ldots,6\}$) moved to
rank $j$ in round $t+1$.



We see in Table \ref{tab:trans} that keeping the first rank position
for two consecutive rounds happened to a slightly larger extent in the \baseShort competitions ($0.68$) than in the \divShort competitions
($0.63$). (Refer to the cell (1,1) in the tables.)  We also see that
the majority of document transitions from the first rank in the \baseShort
competitions was to the second rank ($0.26$). In contrast, in the
\divShort competitions, the majority of document transitions from the first
rank was to the third rank ($0.23$). The contrast in the latter two findings can be
explained as follows. We re-affirmed theoretically in Section
\ref{sec:model} previous findings \cite{raifer_information_2017} that in the quest for the first rank position players mimic documents highly ranked in the past for the query. Hence, if document $d_1$ at the first rank is replaced in the next round with a highly similar document $d_2$, then due to the MMR diversification algorithm in the \divShort competitions $d_1$'s retrieval score is highly likely to be quite penalized; hence, the rank drop. In contrast, in the \baseShort competition $d_1$'s retrieval score is not penalized for its similarity with $d_2$.


We observed that about $13\%$ of the documents in each of the \baseShort and \divShort competitions were identical to the initial document provided to the students before the first round. These documents were typically submitted by players who consistently held the lowest rank position throughout most of the repeated game and were less engaged or  dropped out of the competition. Thus, in what follows, we exclude these documents from the analysis to focus on active players who were engaged in the competition.
\newcommand{\figWidth}{1in}
\newcommand{\figHeight}{1in}
\newcommand{\redSpace}{-.1in}
\begin{figure}[t]
\centering
  \begin{tabular}{ccc}
    \multicolumn{3}{c}{\textbf{Query dependent features}} \\
    \includegraphics[width=\figWidth,height=\figHeight]{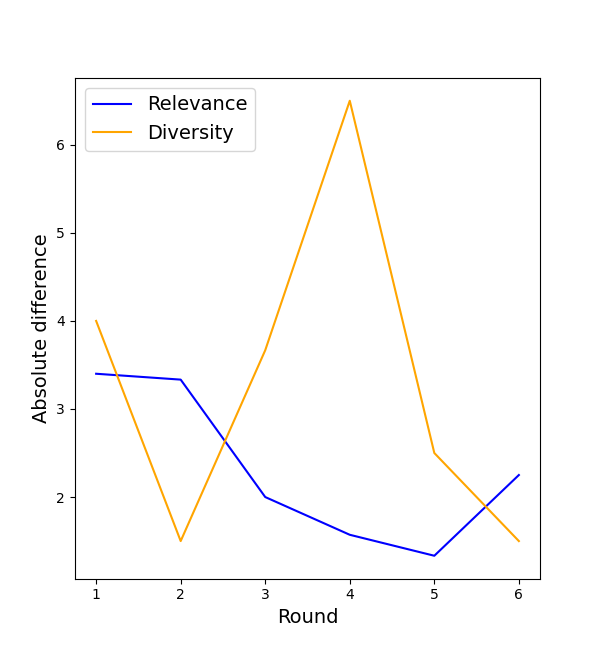} & \hspace*{\redSpace} \includegraphics[width=\figWidth,height=\figHeight]{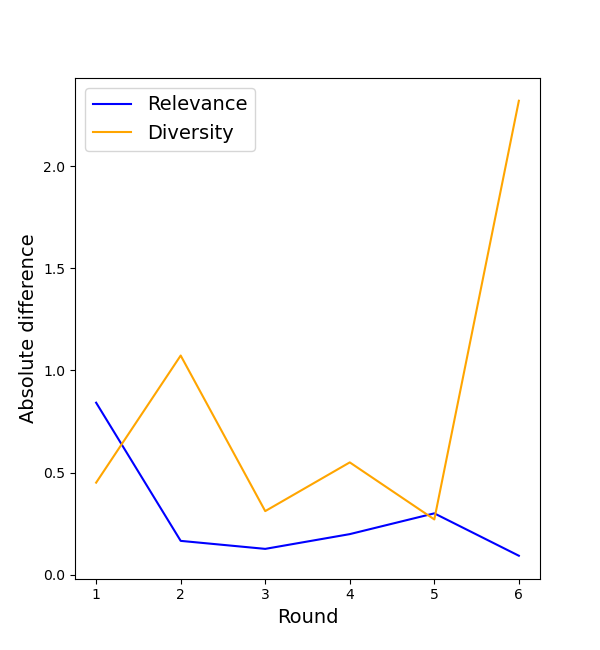} & \hspace*{\redSpace} \includegraphics[width=\figWidth,height=\figHeight]{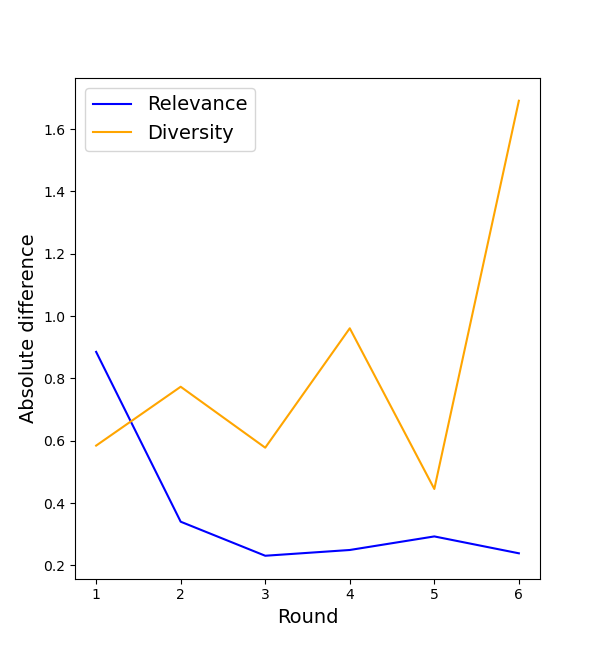} \\
    TF & BM25 & LM.DIR \\
    \multicolumn{3}{c}{\textbf{Query independent features}} \\
    \includegraphics[width=\figWidth,height=\figHeight]{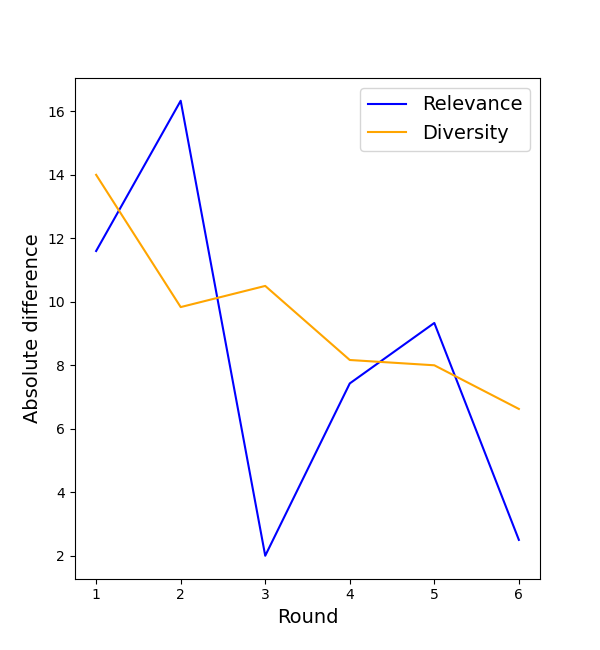} & \hspace*{\redSpace} \includegraphics[width=\figWidth,height=\figHeight]{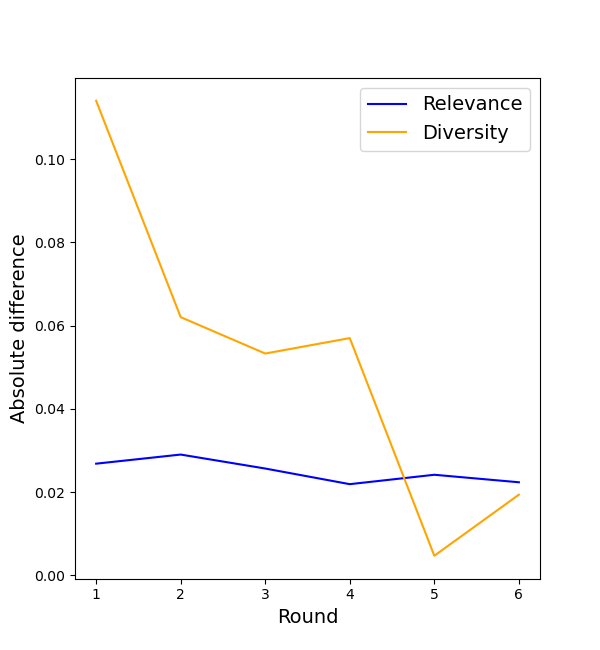} & \hspace*{\redSpace} \includegraphics[width=\figWidth,height=\figHeight]{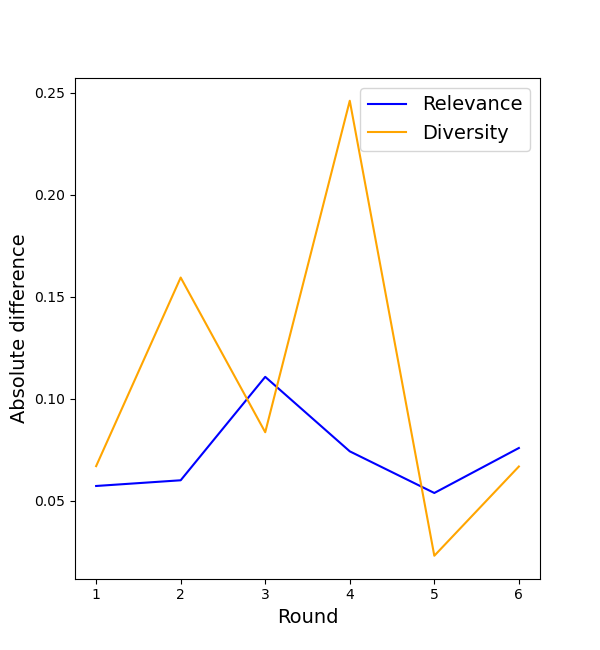} \\
    Length & StopwordRatio & Entropy \\
  \end{tabular}
  \caption{\label{fig_feature_strategies} Average absolute difference of feature values of winner documents in rounds $i$ ($W_i$) and $i+1$ ($W_{i+1}$).}
\end{figure}
\newcommand{\figWidthSecond}{1.6in}
\newcommand{\figHeightSecond}{1.2in}
\newcommand{\figWidthThird}{\figWidthSecond}
\newcommand{\figHeightThird}{\figHeightSecond}
\newcommand{\redSpaceTwo}{-.2in}
\begin{figure}[t]
  \begin{tabular}{cc}
    \includegraphics[width=\figWidthThird,height=\figHeightThird]{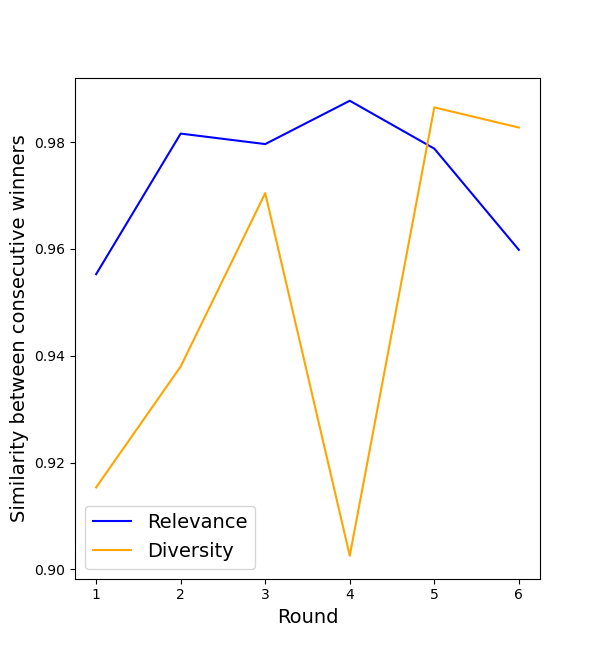}
    & 
    \includegraphics[width=\figWidthThird,height=\figHeightThird]{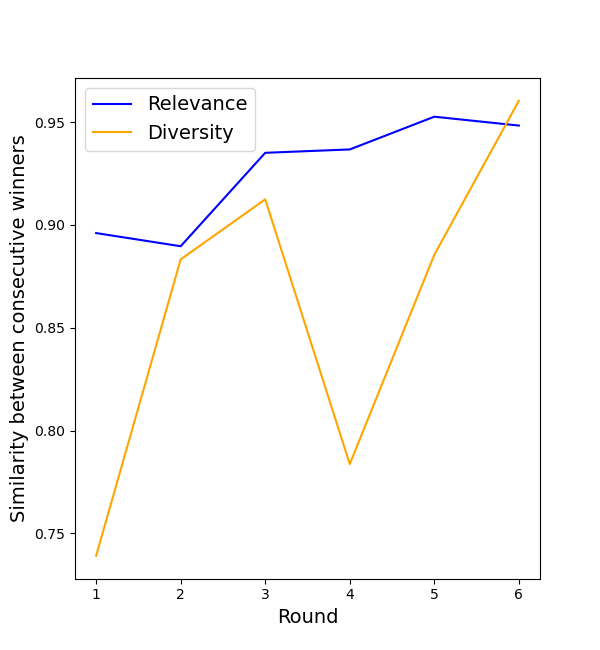}
    \\
    {E5} & {SBERT} \\
    \includegraphics[width=\figWidthThird,height=\figHeightThird]{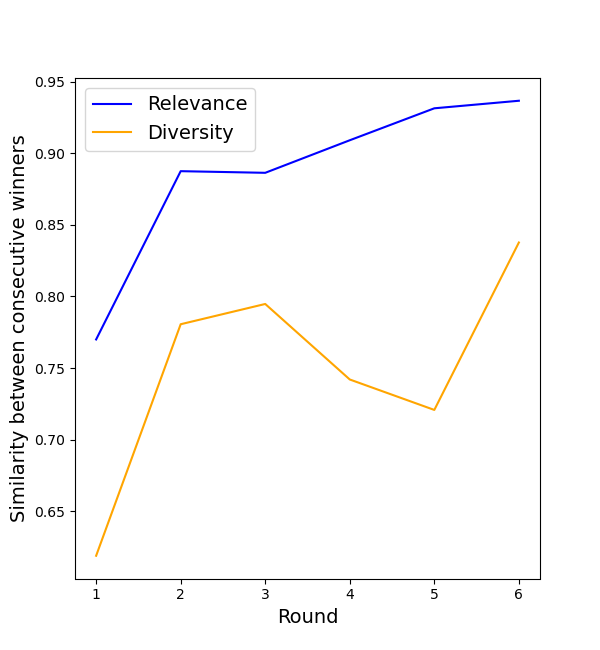}
    & 
    \includegraphics[width=\figWidthThird,height=\figHeightThird]{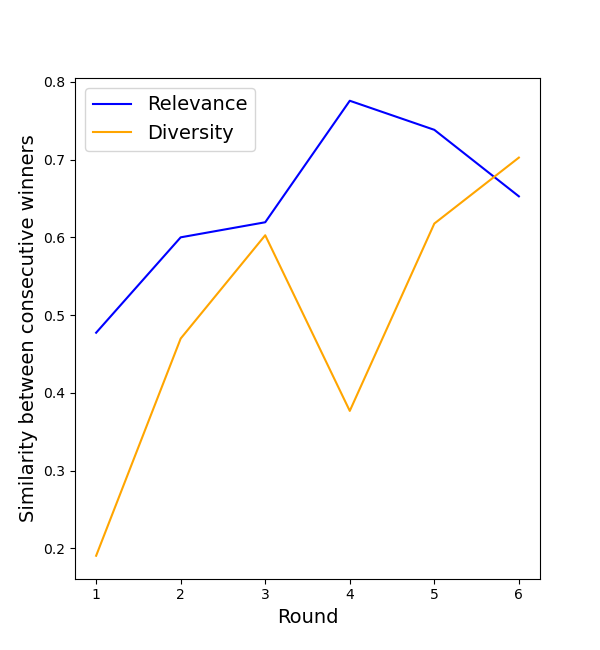}\\
                    {TF.IDF} & {Jaccard} \\
                      \end{tabular}
  \caption{\label{fig:cons_embeddings} The average (over queries) similarity between consecutive winner documents ($W_i$ and $W_{i+1}$).}
  \end{figure}

\subsubsection{Document modifications}\label{sec:doc_mod}
Inspired by the analysis performed for ranking competitions where
relevance was the sole criterion for ranking
\cite{raifer_information_2017}, we now turn to study document
modifications in our diversity-based setting. Specifically, we
analyzed changes in feature values of winner documents (i.e., the
highest ranked) between consecutive rounds; $W_i$ and $W_{i+1}$ are the
winner documents in rounds $i$ and $i+1$, respectively. Our analysis
focuses on cases where $W_i$ and $W_{i+1}$ are produced by different
players as it was observed in prior work
\cite{raifer_information_2017} that players who win a round are
unlikely to substantially change their document for the next round.


As Raifer et al. \cite{raifer_information_2017}, we use a few representative query independent and query dependent
(i.e., prior relevance estimates) features, most of which were used in Microsoft's
learning-to-rank datasets\footnote{\url{www.research.microsoft.com/en-us/projects/mslr}}. The query-dependent features are (i) \firstmention{TF}: the sum of tf
values of query terms in a document, (ii) \firstmention{BM25}: the
Okapi {BM25} retrieval score of the document, and (iii)
\firstmention{LM.DIR}: the query likelihood score of a document where
document language models are Dirichlet smoothed with smoothing
parameter set to $1000$ \cite{Zhai+Lafferty:01a}. The query-independent
features are: (iv) \firstmention{Length}: document length, (v)
\firstmention{StopwordRatio}: the ratio of stopwords to non stopwords
in the document; the INQUERY stopword list was used
\cite{Allan+al:00a}; high presence of stopwords was shown to be
correlated with relevance in Web retrieval \cite{Bendersky+al:11a}, and (vi) \firstmention{Entropy}: the entropy of the unsmoothed unigram
maximum likelihood estimate induced from the document; higher entropy
implies content diversity which can indicate relevance 
\cite{Kurland+Lee:05a}.

Figure \ref{fig_feature_strategies} presents the absolute difference of the feature values of $W_i$ and $W_{i+1}$ (i.e., two consecutive winner documents) for the \Baseline (\baseShort) and \Diversity (\divShort) competitions.
We see that the curve for \divShort for all features, and for almost
all rounds, is in most cases higher than the curve for
\baseShort. Indeed, the average over rounds of the absolute
difference of the feature value for $W_i$ and $W_{i+1}$ for features
(i) - (vi) for the \baseShort (\divShort) competitions is $2.17$
($3.24$), $0.3$ ($1$), $0.37$ ($0.95$), $8.03$ ($9.39$), $0.02$
($0.05$) and $0.07$ ($0.12$), respectively.

As an additional analysis of the relation between consecutive winner
documents ($W_i$ and $W_{i+1}$) we present in Figure
\ref{fig:cons_embeddings} their similarity (averaged over
queries). The similarity measures are: (i) the cosine between E5
\cite{wang_text_2024} document embeddings\footnote{Recall that ranking
  in the competitions is based on E5 embedding.}, (ii) the cosine
between sentence-bert (SBERT) document embeddings
\cite{reimers_sentence-bert_2019}, (iii) the cosine between TF.IDF
document vectors\footnote{To induce robust IDF values, we used the
  competition corpus and TREC's ClueWeb09 corpus
  (\url{https://lemurproject.org/clueweb09.php}). Documents were
  Krovetz stemmed.}, and (iv) Jaccard. We see that the similarity of consecutive
winner documents in the \divShort competitions is lower than that for the \baseShort competitions in almost all rounds and for all similarity measures. The average over rounds for the E5, SBERT, TF.IDF and Jaccard similarity for the \baseShort (\divShort) competitions is: $0.97$ ($0.95$), $0.93$ ($0.87$), $0.89$ ($0.76$), $0.66$ ($0.55$), respectively. This further demonstrates that the similarity between a winner and previous winner documents in the \baseShort competitions is higher than in the \divShort competitions.

Hence, we conclude that to
become a winner, a player modified her document with respect to the
previous winner document to a larger extent in the \divShort (\Diversity)
competition than in the \baseShort (\Baseline) competition. This finding echoes
our theoretical results from Section \ref{sec:model}: when search
results diversification is applied, the ``mimicking the winner''
phenomenon (i.e., making documents similar to those highly ranked in
the past) \cite{raifer_information_2017} is
ameliorated. Interestingly, the players (students) were not actually
informed that diversification was applied. Still, there is past
evidence in ranking competitions \cite{goren_driving_2021} that
players manage to make correct  subtle observations about
properties of the undisclosed ranking function.

\subsection{Inter-document similarities in ranked lists}
\label{sec:interDoc}



\begin{table}[t]
  \caption{\label{table_macro} The average over rounds and queries of
    the mean and minimum inter-document similarity in a ranked
    list for the \baseShort and \divShort competitions. '*' marks a
    statistically significant difference between \divShort and \baseShort.}  \scriptsize
    \begin{tabular}{l|cccc}
&   {E5 (\baseShort, \divShort)} & {SBERT (\baseShort, \divShort)} & {TF.IDF (\baseShort, \divShort)} & {Jaccard (\baseShort, \divShort)} \\ \toprule
Mean similarity      & $0.94$, $0.9^{*}$ & $0.85$, $0.77^{*}$ & $0.74$, $0.61^{*}$ & $0.44$, $0.35^{*}$ \\
Min similarity& $0.91$, $0.85^{*}$ & $0.8$, $0.67^{*}$ & $0.65$, $0.46^{*}$ & $0.32$, $0.22^{*}$ \\ \bottomrule
\end{tabular}

\end{table}

\begin{figure}[t]
  \begin{tabular}{cc}
    \includegraphics[width=\figWidthThird,height=\figHeightThird]{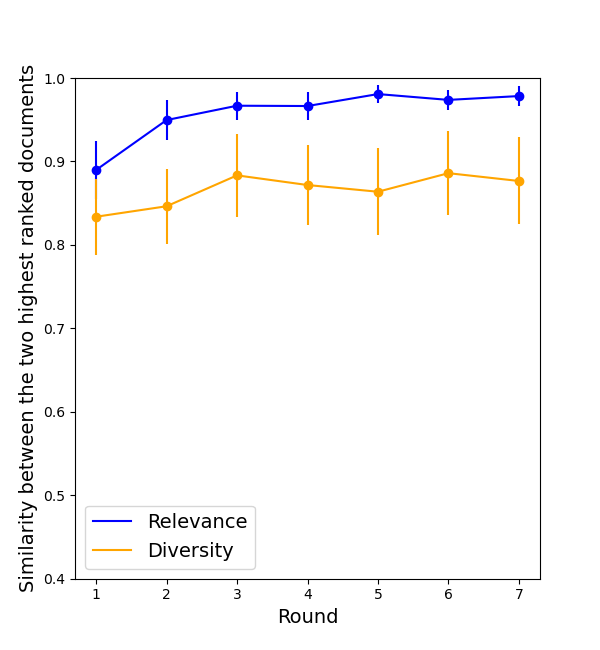}
    & 
    \includegraphics[width=\figWidthThird,height=\figHeightThird]{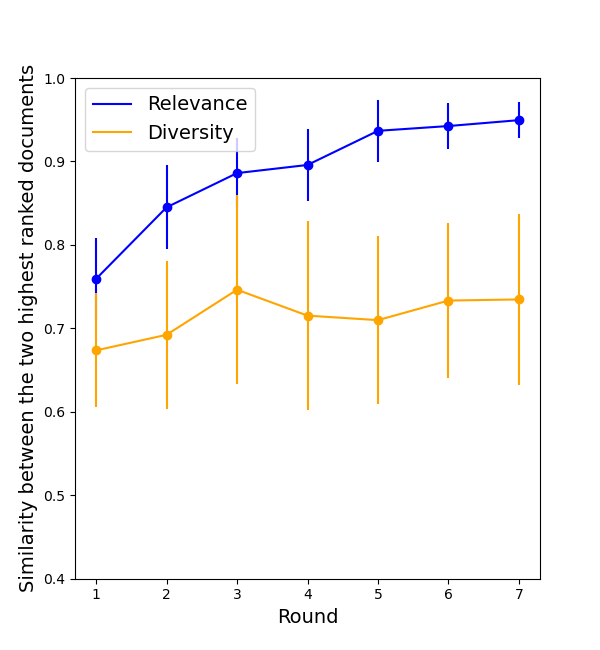}
    \\
    {E5} & {SBERT} \\
    \includegraphics[width=\figWidthThird,height=\figHeightThird]{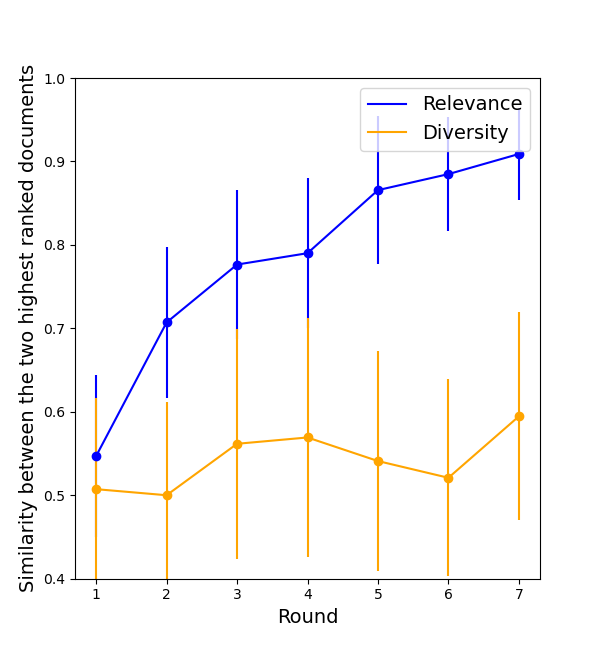}
    & 
    \includegraphics[width=\figWidthThird,height=\figHeightThird]{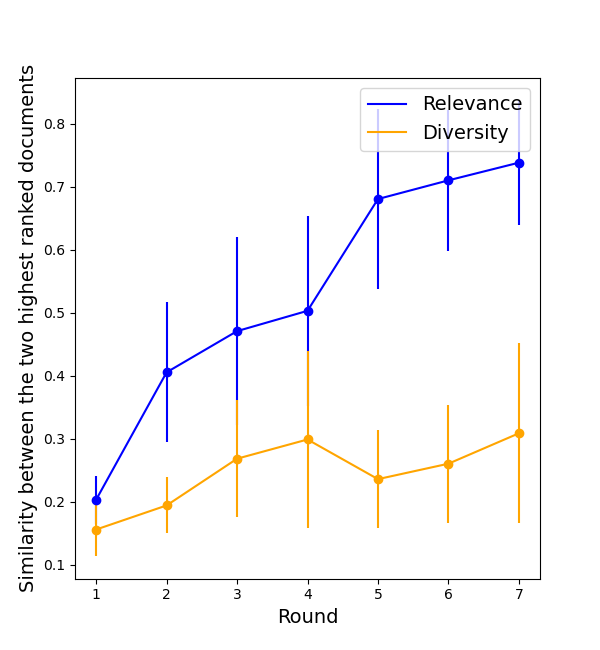}\\
                    {TF.IDF} & {Jaccard} \\
                      \end{tabular}
  \caption{\label{fig:firstSecond} The average (over queries)
    similarity (with confidence intervals) between the two highest
    ranked documents.}
  \end{figure}

We now turn to analyze inter-document similarities in a ranked list.
In Table \ref{table_macro} we report for the \Baseline (\baseShort) and \Diversity (\divShort) competitions the mean and minimum inter-document similarity in a ranked list (averaged over rounds and queries) measured with the four similarity estimates described in Section \ref{sec:doc_mod}. We see that the inter-document similarity
values are in all cases higher --- to a statistically significant
degree --- for the
\baseShort competitions than for the \divShort competitions. This
finding is in accordance with our theoretical results from Section
\ref{sec:model}. That is, we showed that as from a certain point, some
players in competitions with diversity-based ranking aim to secure the
second rank position. To that end, their documents must differ from
the highest ranked documents. By induction, some players will secure
the third place by having their documents differ from the first
two. Hence, inter-document similarities are relatively not high. In contrast, in the \baseShort competitions, as we showed in
Section \ref{sec:model} and as shown in previous work
\cite{raifer_information_2017}, players continuously compete for the
first place by ``mimicking the winner'' which is the min-max regret
equilibrium strategy. This results in documents being quite similar to each other.

Based on the findings just mentioned, and those in Section
\ref{sec:doc_mod} about diversity-based ranking resulting in
ameliorated ``mimicking the winner'' strategy, we conclude that
diversity-based ranking helps to ameliorate the herding effect with
respect to ranking solely based on relevance estimation.

\subsection{Temporal dynamics}
We next turn to explore the changes along the competitions' rounds of
several types of similarities.

\myparagraph{The similarity between the two highest ranked documents}
In Figure \ref{fig:firstSecond} we present the average similarity (across queries), and corresponding confidence intervals, of the similarity between the two highest ranked documents in a list along the competition rounds. We see that the similarity for the \Baseline (\baseShort) competitions is monotonically increasing to a much larger extent than for the \Diversity (\divShort) competitions. Furthermore, the similarities for the \baseShort competitions are consistently (along rounds) statistically significantly higher than those for the \divShort competitions\footnote{To increase the sample size for each comparison group from $15$ (queries per round) to $30$, statistical significance tests were performed on pairs of consecutive rounds.}. These findings are expected: when ranking is solely based on relevance, players who were not ranked first make their documents more similar to those most highly ranked in the past \cite{raifer_information_2017}. In contrast, the MMR-based ranking employed in the \divShort competitions, along with our finding that some players will ``give up'' on winning the competition and will strive to secure the second place (see Section \ref{sec:model}), results in lower similarity between the two highest ranked documents. Thus, we get further empirical support to the fact that diversity-based ranking helps to ameliorate the ``mimicking the winner'' strategy. 

The confidence intervals in Figure \ref{fig:firstSecond} for the similarity between the two highest ranked documents are often (much) larger for the \divShort than for the \baseShort competitions. This finding attests to the transition from competing for the first rank position to competing for the second rank position which emerged in our theoretical analysis in Section \ref{sec:model}.

\begin{figure}[t]  
\begin{tabular}{cc}
        \includegraphics[width=\figWidthSecond,height=\figHeightSecond]{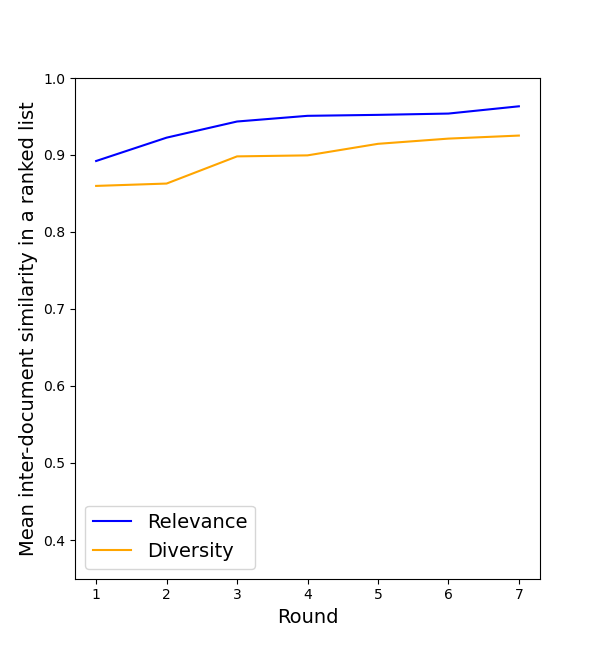}
        &
        \includegraphics[width=\figWidthSecond,height=\figHeightSecond]{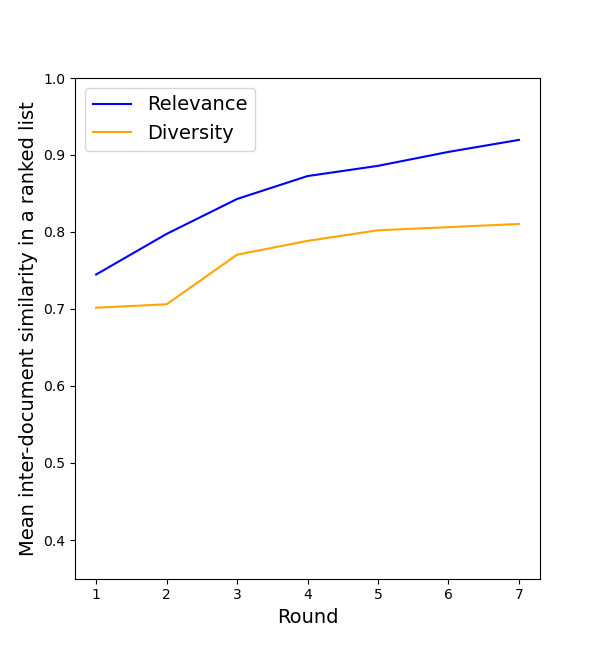} \\
         E5 & SBERT \\
        \includegraphics[width=\figWidthSecond,height=\figHeightSecond]{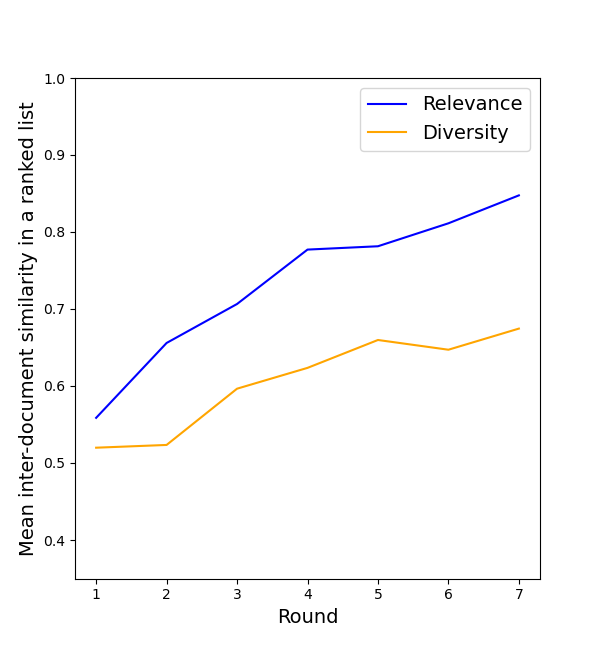}
&
        \includegraphics[width=\figWidthSecond,height=\figHeightSecond]{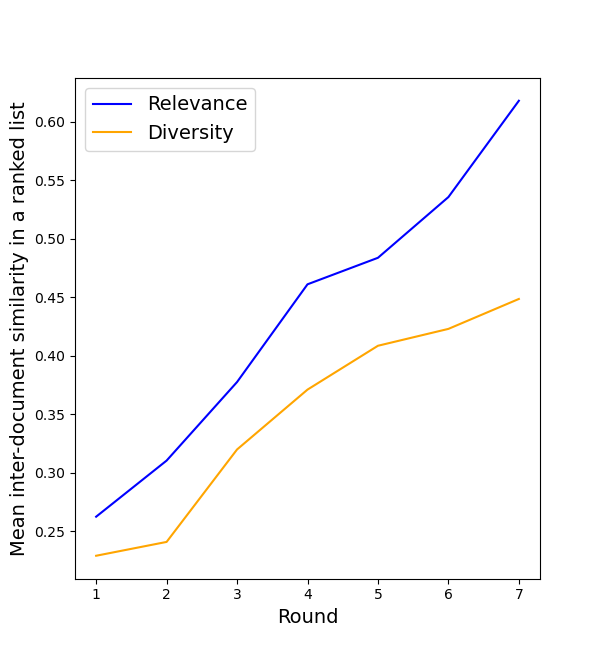}\\
                 TF.IDF &  Jaccard \\
\end{tabular}
\caption{\label{fig:listSimRounds} The average (over queries) mean inter-document similarity in ranked list in a round.}
\end{figure}

\myparagraph{Inter-document similarities in lists}
In Section \ref{sec:interDoc} we studied the inter-document similarities in ranked lists over the entire competition (i.e., averaged over rounds). We now turn to analyze the temporal changes of these similarities along the competition rounds. Figure \ref{fig:listSimRounds} presents the (average over queries) of the mean inter-document similarity in a ranked list per round.

We see in Figure \ref{fig:listSimRounds} that the mean inter-document
similarity in a ranked list for the \Baseline (\baseShort)
competitions is consistently higher than that for the \Diversity
(\divShort) competitions. The differences between the \baseShort and
\divShort competitions are statistically significant (using the same
statistical significance test used above) in all cases for the E5 and
SBERT similarity estimates and in almost all cases for TF.IDF; for
Jaccard, the difference was rarely statistically significant.

These findings about the temporal patterns of inter-document similarities in the ranked lists,
together with the findings above about a reduced ``mimicking the
winner'' phenomenon in \divShort with respect to \baseShort
competitions, lead again to the conclusion that diversity-based ranking
helps to ameliorate to some extent publisher herding.

\begin{figure}[t]
  \begin{tabular}{cc}
    \includegraphics[width=\figWidthSecond,height=\figHeightSecond]{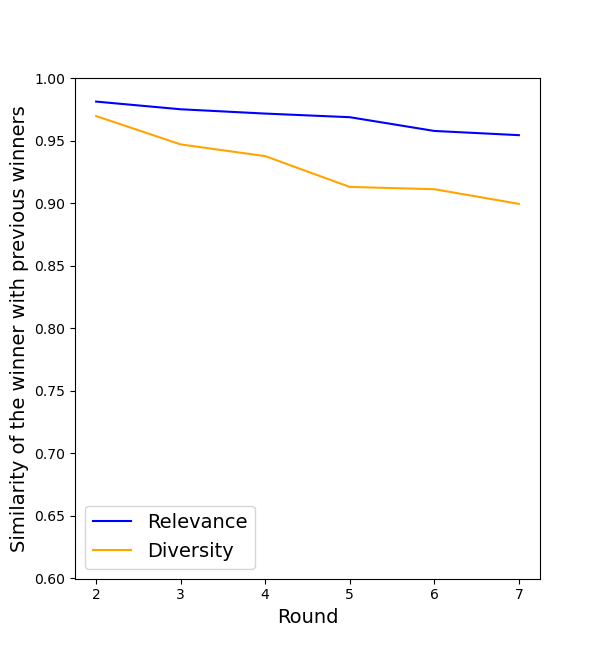} &         \includegraphics[width=\figWidthSecond,height=\figHeightSecond]{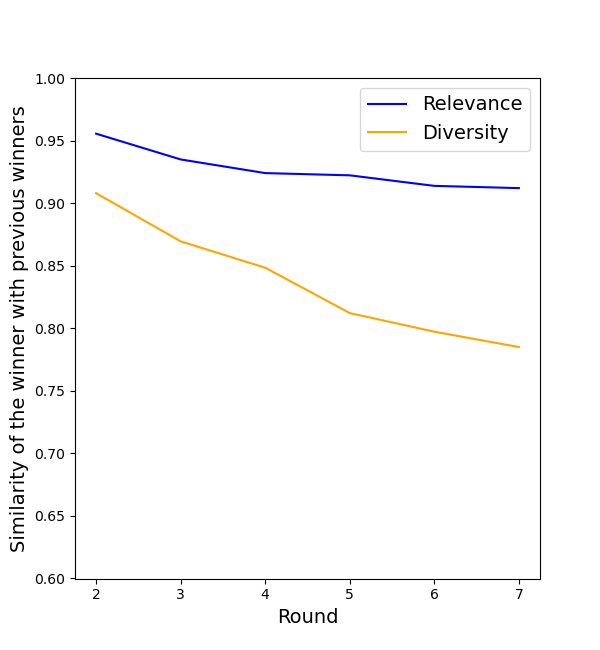} \\
    E5 & SBERT \\
    \includegraphics[width=\figWidthSecond,height=\figHeightSecond]{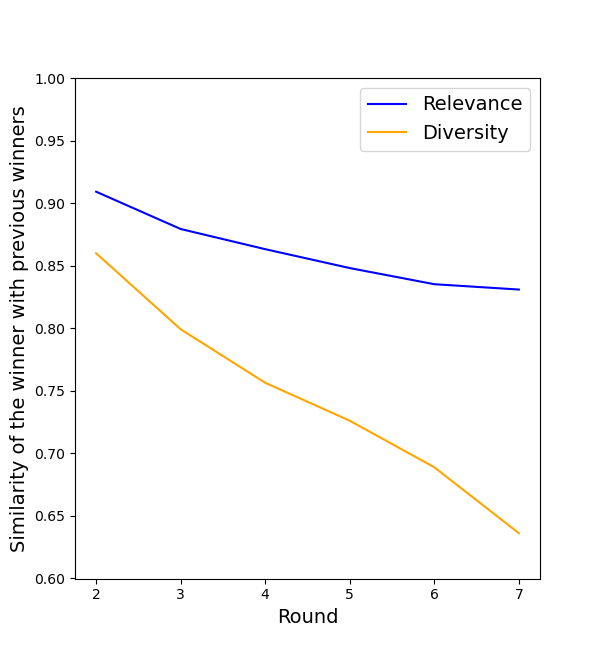} &         \includegraphics[width=\figWidthSecond,height=\figHeightSecond]{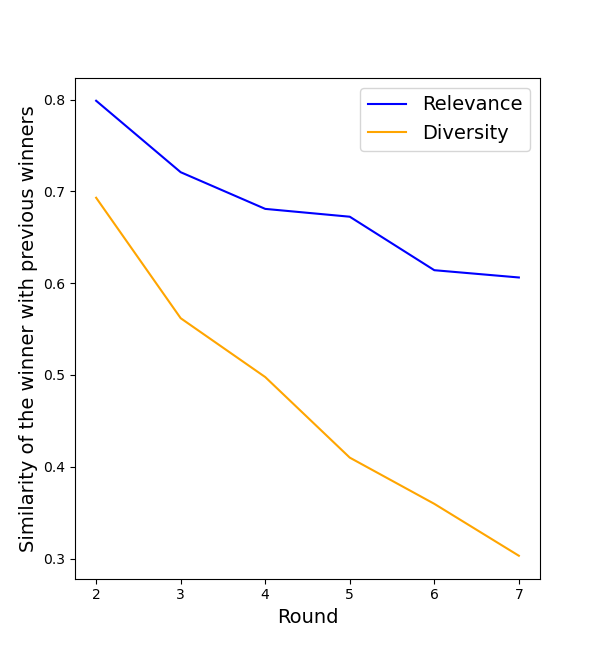} \\
    TD.IDF & Jaccard \\
  \end{tabular}
  \caption{\label{fig:firstSim} The minimum (over queries) similarity between the highest ranked document in round $i$ (``winner'') and the winners in each of the rounds $1,\ldots,i-1$.}
  \end{figure}

\myparagraph{Dynamics of winners} Figure \ref{fig:firstSim} shows the minimum
inter-document similarity between the highest ranked document in a round (``winner'')
and all winners of previous rounds. While the similarity decreases for
both types of competitions (\baseShort and \divShort), the decrease is
much more substantial for the \divShort than for the \baseShort
competitions. This substantial change in the content of winner
documents for the \Diversity (\divShort) competitions can be explained using the
finding in Table \ref{tab:trans}: winners who lost the first place
were much more likely to move to the third rank position than to the
second. Hence, presumably to avoid this rank drop, winners were
changing their documents so as to maintain the first rank position. In addition, Table \ref{tab:trans} shows that the move to the first rank position was in most cases from the third rank position which was populated by documents quite dissimilar to the first two documents. Given the ``risk'' in becoming too similar to these documents, winning was achieved using documents dissimilar to those ranked above them.



\section{Conclusions and Future Work}
In competitive search settings \cite{kurland_competitive_2022},
publishers of documents are incentivized to have them highly
ranked. As a result, the publishers respond to induced rankings by
modifying their documents with the goal of improving their future
ranking.

Previous work on competitive search focused on ranking
functions based solely on relevance estimation. We present the first
theoretical and empirical analysis of a competitive search setting
where search-results diversification is applied.

Our motivation to study the effects of diversity-based ranking is
rooted in previous findings about a prevalent strategy of publishers:
mimicking content in documents most highly ranked in the past
\cite{raifer_information_2017}. This strategy was shown to lead to
herding of publishers with unwarranted corpus effects
\cite{goren_driving_2021}; e.g., reduced topical diversity in the
corpus. The main research question that naturally emerges is whether diversity-based ranking can help to reduce the extent to which the content mimicking strategy is applied, and consequently to ameliorate herding.

We presented a game theoretic analysis of the competitive search setting with diversity-based ranking. We showed that there is a min-max regret equilibrium which means stability. We also showed that some publishers will focus on trying to secure the second rank position, and to this end, will have to make their documents less similar to the highest ranked ones. As a result, the mimicking strategy becomes less prevalent and herding is accordingly ameliorated.

For empirical analysis, we organized ranking competitions between students where the ranking
function either included a search-results diversification mechanism or
not. We found that with diversity-based ranking, the
mimicking strategy was less prevalent than when no diversification was
applied. Together with the overall increased content diversity we provided empirical support to the fact that diversity-based ranking helps to ameliorate publisher herding.

As in almost all previous work on competitive search, we assumed that
a publisher modifies her document to improve ranking for a single
query. There is only one report we are aware of on competitive search with publishers
competing for multiple queries representing the same information need \cite{nachimovsky_ranking-incentivized_2024}. Ranking was based solely on relevance estimation. Accordingly, for future work we plan to analyze the multiple-queries setting where the retrieval method applies results diversification.

\balance
\bibliographystyle{ACM-Reference-Format}

\begin{thebibliography}{39}


\ifx \showCODEN    \undefined \def \showCODEN     #1{\unskip}     \fi
\ifx \showDOI      \undefined \def \showDOI       #1{#1}\fi
\ifx \showISBNx    \undefined \def \showISBNx     #1{\unskip}     \fi
\ifx \showISBNxiii \undefined \def \showISBNxiii  #1{\unskip}     \fi
\ifx \showISSN     \undefined \def \showISSN      #1{\unskip}     \fi
\ifx \showLCCN     \undefined \def \showLCCN      #1{\unskip}     \fi
\ifx \shownote     \undefined \def \shownote      #1{#1}          \fi
\ifx \showarticletitle \undefined \def \showarticletitle #1{#1}   \fi
\ifx \showURL      \undefined \def \showURL       {\relax}        \fi
\providecommand\bibfield[2]{#2}
\providecommand\bibinfo[2]{#2}
\providecommand\natexlab[1]{#1}
\providecommand\showeprint[2][]{arXiv:#2}

\bibitem[noa(2024)]%
        {noauthor_introducing_2024}
 \bibinfo{year}{2024}\natexlab{}.
\newblock \bibinfo{title}{Introducing {Connect} by {CloudResearch}: {Advancing} {Online} {Participant} {Recruitment} in the {Digital} {Age} {\textbar} {Request} {PDF}}.
\newblock
\newblock
\urldef\tempurl%
\url{https://doi.org/10.31234/osf.io/ksgyr}
\showDOI{\tempurl}


\bibitem[Allan et~al\mbox{.}(2000)]%
        {Allan+al:00a}
\bibfield{author}{\bibinfo{person}{James Allan}, \bibinfo{person}{Margaret~E. Connell}, \bibinfo{person}{W.~Bruce Croft}, \bibinfo{person}{Fang-Fang Feng}, \bibinfo{person}{David Fisher}, {and} \bibinfo{person}{Xiaoyan Li}.} \bibinfo{year}{2000}\natexlab{}.
\newblock \showarticletitle{{INQUERY} and {TREC}-9}. In \bibinfo{booktitle}{\emph{Proceedings of {TREC}}}. \bibinfo{pages}{551--562}.
\newblock


\bibitem[Aumann et~al\mbox{.}(1995)]%
        {aumann1995repeated}
\bibfield{author}{\bibinfo{person}{Robert~J Aumann}, \bibinfo{person}{Michael Maschler}, {and} \bibinfo{person}{Richard~E Stearns}.} \bibinfo{year}{1995}\natexlab{}.
\newblock \bibinfo{booktitle}{\emph{Repeated games with incomplete information}}.
\newblock \bibinfo{publisher}{MIT press}.
\newblock


\bibitem[Banerjee(1992)]%
        {Banerjee}
\bibfield{author}{\bibinfo{person}{Banerjee}.} \bibinfo{year}{1992}\natexlab{}.
\newblock \showarticletitle{A simple model of herd behavior}.
\newblock \bibinfo{journal}{\emph{The Quarterly Journal of Economics}}  \bibinfo{volume}{107} (\bibinfo{year}{1992}), \bibinfo{pages}{797--817}.
\newblock


\bibitem[Ben{-}Basat et~al\mbox{.}(2017)]%
        {Basat+al:17a}
\bibfield{author}{\bibinfo{person}{Ran Ben{-}Basat}, \bibinfo{person}{Moshe Tennenholtz}, {and} \bibinfo{person}{Oren Kurland}.} \bibinfo{year}{2017}\natexlab{}.
\newblock \showarticletitle{A Game Theoretic Analysis of the Adversarial Retrieval Setting}.
\newblock \bibinfo{journal}{\emph{J. Artif. Intell. Res.}}  \bibinfo{volume}{60} (\bibinfo{year}{2017}), \bibinfo{pages}{1127--1164}.
\newblock


\bibitem[Bendersky et~al\mbox{.}(2011)]%
        {Bendersky+al:11a}
\bibfield{author}{\bibinfo{person}{Michael Bendersky}, \bibinfo{person}{W.~Bruce Croft}, {and} \bibinfo{person}{Yanlei Diao}.} \bibinfo{year}{2011}\natexlab{}.
\newblock \showarticletitle{Quality-biased ranking of web documents}. In \bibinfo{booktitle}{\emph{Proceedings of WSDM}}. \bibinfo{pages}{95--104}.
\newblock


\bibitem[Bikhchandani et~al\mbox{.}(1992)]%
        {Bikhchandani}
\bibfield{author}{\bibinfo{person}{S. Bikhchandani}, \bibinfo{person}{D. Hirshleifer}, {and} \bibinfo{person}{I. Welch}.} \bibinfo{year}{1992}\natexlab{}.
\newblock \showarticletitle{A theory of fads, fashion, custom and cultural change as information cascade}.
\newblock \bibinfo{journal}{\emph{The Journal of Political Economy}}  \bibinfo{volume}{100} (\bibinfo{year}{1992}), \bibinfo{pages}{992--1026}.
\newblock


\bibitem[Carbonell and Goldstein(1998)]%
        {carbonell_use_1999}
\bibfield{author}{\bibinfo{person}{Jaime~G. Carbonell} {and} \bibinfo{person}{Jade Goldstein}.} \bibinfo{year}{1998}\natexlab{}.
\newblock \showarticletitle{The Use of MMR, Diversity-Based Reranking for Reordering Documents and Producing Summaries}. In \bibinfo{booktitle}{\emph{Proceedings of SIGIR}}. \bibinfo{pages}{335--336}.
\newblock


\bibitem[Castillo and Davison(2010)]%
        {Castillo+Davison:10a}
\bibfield{author}{\bibinfo{person}{Carlos Castillo} {and} \bibinfo{person}{Brian~D. Davison}.} \bibinfo{year}{2010}\natexlab{}.
\newblock \showarticletitle{Adversarial Web Search}.
\newblock \bibinfo{journal}{\emph{Foundations and Trends in Information Retrieval}} \bibinfo{volume}{4}, \bibinfo{number}{5} (\bibinfo{year}{2010}), \bibinfo{pages}{377--486}.
\newblock


\bibitem[Chen et~al\mbox{.}(2023)]%
        {Xuanang+al:23a}
\bibfield{author}{\bibinfo{person}{Xuanang Chen}, \bibinfo{person}{Ben He}, \bibinfo{person}{Zheng Ye}, \bibinfo{person}{Le Sun}, {and} \bibinfo{person}{Yingfei Sun}.} \bibinfo{year}{2023}\natexlab{}.
\newblock \showarticletitle{Towards Imperceptible Document Manipulations against Neural Ranking Models}. In \bibinfo{booktitle}{\emph{Findings of the Association for Computational Linguistics: ACL 2023}}.
\newblock


\bibitem[Eilat and Rosenfeld(2023)]%
        {eilat_performative_2023}
\bibfield{author}{\bibinfo{person}{Itay Eilat} {and} \bibinfo{person}{Nir Rosenfeld}.} \bibinfo{year}{2023}\natexlab{}.
\newblock \bibinfo{title}{Performative {Recommendation}: {Diversifying} {Content} via {Strategic} {Incentives}}.
\newblock
\newblock
\urldef\tempurl%
\url{http://arxiv.org/abs/2302.04336}
\showURL{%
\tempurl}
\newblock
\shownote{arXiv:2302.04336 [cs]}.


\bibitem[Goren et~al\mbox{.}(2020)]%
        {goren_ranking-incentivized_2020}
\bibfield{author}{\bibinfo{person}{Gregory Goren}, \bibinfo{person}{Oren Kurland}, \bibinfo{person}{Moshe Tennenholtz}, {and} \bibinfo{person}{Fiana Raiber}.} \bibinfo{year}{2020}\natexlab{}.
\newblock \showarticletitle{Ranking-{Incentivized} {Quality} {Preserving} {Content} {Modification}}. In \bibinfo{booktitle}{\emph{Proceedings of SIGIR}}. \bibinfo{address}{Virtual Event China}, \bibinfo{pages}{259--268}.
\newblock


\bibitem[Goren et~al\mbox{.}(2021)]%
        {goren_driving_2021}
\bibfield{author}{\bibinfo{person}{Gregory Goren}, \bibinfo{person}{Oren Kurland}, \bibinfo{person}{Moshe Tennenholtz}, {and} \bibinfo{person}{Fiana Raiber}.} \bibinfo{year}{2021}\natexlab{}.
\newblock \showarticletitle{Driving the {Herd}: {Search} {Engines} as {Content} {Influencers}}. In \bibinfo{booktitle}{\emph{Proceedings of CIKM}}. \bibinfo{address}{Virtual Event Queensland Australia}, \bibinfo{pages}{586--595}.
\newblock


\bibitem[Gy{\"o}ngyi and Garcia-Molina(2005)]%
        {Gyongyi+Molina:05a}
\bibfield{author}{\bibinfo{person}{Zolt{\'a}n Gy{\"o}ngyi} {and} \bibinfo{person}{Hector Garcia-Molina}.} \bibinfo{year}{2005}\natexlab{}.
\newblock \showarticletitle{Web Spam Taxonomy}. In \bibinfo{booktitle}{\emph{Proceedings of AIRWeb 2005}}. \bibinfo{pages}{39--47}.
\newblock


\bibitem[Hyafil and Boutilier(2012)]%
        {Hyafil+Boutilier:12a}
\bibfield{author}{\bibinfo{person}{Nathanael Hyafil} {and} \bibinfo{person}{Craig Boutilier}.} \bibinfo{year}{2012}\natexlab{}.
\newblock \showarticletitle{Regret Minimizing Equilibria and Mechanisms for Games with Strict Type Uncertainty}.
\newblock \bibinfo{journal}{\emph{CoRR}}  \bibinfo{volume}{abs/1207.4147} (\bibinfo{year}{2012}).
\newblock


\bibitem[Jenkins et~al\mbox{.}(2020)]%
        {Jenkins+al:20a}
\bibfield{author}{\bibinfo{person}{Porter Jenkins}, \bibinfo{person}{Jennifer Zhao}, \bibinfo{person}{Heath Vinicombe}, \bibinfo{person}{Anant Subramanian}, \bibinfo{person}{Arun Prasad}, \bibinfo{person}{Atillia Dobi}, \bibinfo{person}{Eileen Li}, {and} \bibinfo{person}{Yunsong Guo}.} \bibinfo{year}{2020}\natexlab{}.
\newblock \showarticletitle{Natural Language Annotations for Search Engine Optimization}. In \bibinfo{booktitle}{\emph{Proceedings of The Web Conference}}. \bibinfo{pages}{2856--2862}.
\newblock


\bibitem[Joachims et~al\mbox{.}(2005)]%
        {Joachims+al:05a}
\bibfield{author}{\bibinfo{person}{Thorsten Joachims}, \bibinfo{person}{Laura Granka}, \bibinfo{person}{Bing Pan}, \bibinfo{person}{Helene Hembrooke}, {and} \bibinfo{person}{Geri Gay}.} \bibinfo{year}{2005}\natexlab{}.
\newblock \showarticletitle{Accurately interpreting clickthrough data as implicit feedback}. In \bibinfo{booktitle}{\emph{Proceedings of SIGIR}}. \bibinfo{pages}{154--161}.
\newblock


\bibitem[Kurland and Lee(2005)]%
        {Kurland+Lee:05a}
\bibfield{author}{\bibinfo{person}{Oren Kurland} {and} \bibinfo{person}{Lillian Lee}.} \bibinfo{year}{2005}\natexlab{}.
\newblock \showarticletitle{{PageRank} without hyperlinks: Structural re-ranking using links induced by language models}. In \bibinfo{booktitle}{\emph{Proceedings of SIGIR}}. \bibinfo{pages}{306--313}.
\newblock


\bibitem[Kurland and Tennenholtz(2022)]%
        {kurland_competitive_2022}
\bibfield{author}{\bibinfo{person}{Oren Kurland} {and} \bibinfo{person}{Moshe Tennenholtz}.} \bibinfo{year}{2022}\natexlab{}.
\newblock \showarticletitle{Competitive {Search}}. In \bibinfo{booktitle}{\emph{Proceedings of SIGIR}}. \bibinfo{pages}{2838--2849}.
\newblock


\bibitem[Liu(2011)]%
        {Liu:11a}
\bibfield{author}{\bibinfo{person}{Tie-Yan Liu}.} \bibinfo{year}{2011}\natexlab{}.
\newblock \bibinfo{booktitle}{\emph{Learning to Rank for Information Retrieval}}.
\newblock \bibinfo{publisher}{Springer}. I--XVII, 1--285 pages.
\newblock
\showISBNx{978-3-642-14266-6}


\bibitem[Liu et~al\mbox{.}(2023)]%
        {Liu+al:23a}
\bibfield{author}{\bibinfo{person}{Yu{-}An Liu}, \bibinfo{person}{Ruqing Zhang}, \bibinfo{person}{Jiafeng Guo}, \bibinfo{person}{Maarten de Rijke}, \bibinfo{person}{Wei Chen}, \bibinfo{person}{Yixing Fan}, {and} \bibinfo{person}{Xueqi Cheng}.} \bibinfo{year}{2023}\natexlab{}.
\newblock \showarticletitle{Topic-oriented Adversarial Attacks against Black-box Neural Ranking Models}. In \bibinfo{booktitle}{\emph{Proceedings of SIGIR}}. \bibinfo{pages}{1700--1709}.
\newblock


\bibitem[Liu et~al\mbox{.}(2024)]%
        {Liu+al:24a}
\bibfield{author}{\bibinfo{person}{Yu-An Liu}, \bibinfo{person}{Ruqing Zhang}, \bibinfo{person}{Jiafeng Guo}, \bibinfo{person}{Maarten de Rijke}, \bibinfo{person}{Yixing Fan}, {and} \bibinfo{person}{Xueqi Cheng}.} \bibinfo{year}{2024}\natexlab{}.
\newblock \bibinfo{title}{Robust Neural Information Retrieval: An Adversarial and Out-of-distribution Perspective}.
\newblock
\newblock


\bibitem[Mitra and Craswell(2018)]%
        {mitra_introduction_2018}
\bibfield{author}{\bibinfo{person}{Bhaskar Mitra} {and} \bibinfo{person}{Nick Craswell}.} \bibinfo{year}{2018}\natexlab{}.
\newblock \showarticletitle{An {Introduction} to {Neural} {Information} {Retrieval} t}.
\newblock \bibinfo{journal}{\emph{Foundations and Trends in Information Retrieval}} \bibinfo{volume}{13}, \bibinfo{number}{1} (\bibinfo{year}{2018}), \bibinfo{pages}{1--126}.
\newblock
\showISSN{1554-0669, 1554-0677}


\bibitem[Nachimovsky et~al\mbox{.}(2024)]%
        {nachimovsky_ranking-incentivized_2024}
\bibfield{author}{\bibinfo{person}{Haya Nachimovsky}, \bibinfo{person}{Moshe Tennenholtz}, \bibinfo{person}{Fiana Raiber}, {and} \bibinfo{person}{Oren Kurland}.} \bibinfo{year}{2024}\natexlab{}.
\newblock \showarticletitle{Ranking-{Incentivized} {Document} {Manipulations} for {Multiple} {Queries}}. In \bibinfo{booktitle}{\emph{Proceedings of ICTIR}}. \bibinfo{pages}{61--70}.
\newblock


\bibitem[Raifer et~al\mbox{.}(2017)]%
        {raifer_information_2017}
\bibfield{author}{\bibinfo{person}{Nimrod Raifer}, \bibinfo{person}{Fiana Raiber}, \bibinfo{person}{Moshe Tennenholtz}, {and} \bibinfo{person}{Oren Kurland}.} \bibinfo{year}{2017}\natexlab{}.
\newblock \showarticletitle{Information {Retrieval} {Meets} {Game} {Theory}: {The} {Ranking} {Competition} {Between} {Documents}' {Authors}}. In \bibinfo{booktitle}{\emph{Proceedings of SIGIR}}. \bibinfo{address}{Shinjuku Tokyo Japan}, \bibinfo{pages}{465--474}.
\newblock


\bibitem[Raval and Verma(2020)]%
        {Raval+Verma:20a}
\bibfield{author}{\bibinfo{person}{Nisarg Raval} {and} \bibinfo{person}{Manisha Verma}.} \bibinfo{year}{2020}\natexlab{}.
\newblock \showarticletitle{One word at a time: adversarial attacks on retrieval models}.
\newblock \bibinfo{journal}{\emph{CoRR}}  \bibinfo{volume}{abs/2008.02197} (\bibinfo{year}{2020}).
\newblock


\bibitem[Reimers and Gurevych(2019)]%
        {reimers_sentence-bert_2019}
\bibfield{author}{\bibinfo{person}{Nils Reimers} {and} \bibinfo{person}{Iryna Gurevych}.} \bibinfo{year}{2019}\natexlab{}.
\newblock \showarticletitle{Sentence-BERT: Sentence Embeddings using Siamese BERT-Networks}. In \bibinfo{booktitle}{\emph{Proceedings {EMNLP-IJCNLP}}}. \bibinfo{pages}{3980--3990}.
\newblock


\bibitem[Robertson(1977)]%
        {Robertson:77a}
\bibfield{author}{\bibinfo{person}{Stephen~E. Robertson}.} \bibinfo{year}{1977}\natexlab{}.
\newblock \showarticletitle{The Probability Ranking Principle in {IR}}.
\newblock \bibinfo{journal}{\emph{Journal of Documentation}} (\bibinfo{year}{1977}), \bibinfo{pages}{294--304}.
\newblock
\newblock
\shownote{Reprinted in K. Sparck Jones and P. Willett (eds), {\it Readings in Information Retrieval}, pp. 281--286, 1997}.


\bibitem[Santos et~al\mbox{.}(2015)]%
        {Santos+al:15a}
\bibfield{author}{\bibinfo{person}{Rodrygo L.~T. Santos}, \bibinfo{person}{Craig MacDonald}, {and} \bibinfo{person}{Iadh Ounis}.} \bibinfo{year}{2015}\natexlab{}.
\newblock \showarticletitle{Search Result Diversification}.
\newblock \bibinfo{journal}{\emph{Found. Trends Inf. Retr.}} \bibinfo{volume}{9}, \bibinfo{number}{1} (\bibinfo{year}{2015}), \bibinfo{pages}{1--90}.
\newblock


\bibitem[Smith and Sorensen(2000)]%
        {SmithSorensen}
\bibfield{author}{\bibinfo{person}{L. Smith} {and} \bibinfo{person}{P. Sorensen}.} \bibinfo{year}{2000}\natexlab{}.
\newblock \showarticletitle{Pathalogical outcomes of observational learning}.
\newblock \bibinfo{journal}{\emph{Econometrica}}  \bibinfo{volume}{68} (\bibinfo{year}{2000}), \bibinfo{pages}{371--398}.
\newblock


\bibitem[Song et~al\mbox{.}(2020)]%
        {Song+al:20a}
\bibfield{author}{\bibinfo{person}{Congzheng Song}, \bibinfo{person}{Alexander~M. Rush}, {and} \bibinfo{person}{Vitaly Shmatikov}.} \bibinfo{year}{2020}\natexlab{}.
\newblock \showarticletitle{Adversarial Semantic Collisions}.
\newblock \bibinfo{journal}{\emph{CoRR}}  \bibinfo{volume}{abs/2011.04743} (\bibinfo{year}{2020}).
\newblock


\bibitem[Song et~al\mbox{.}(2022)]%
        {Song+al:22a}
\bibfield{author}{\bibinfo{person}{Junshuai Song}, \bibinfo{person}{Jiangshan Zhang}, \bibinfo{person}{Jifeng Zhu}, \bibinfo{person}{Mengyun Tang}, {and} \bibinfo{person}{Yong Yang}.} \bibinfo{year}{2022}\natexlab{}.
\newblock \showarticletitle{TRAttack: Text Rewriting Attack Against Text Retrieval}. In \bibinfo{booktitle}{\emph{Proceedings of RepL4NLP@ACL}}. \bibinfo{pages}{191--203}.
\newblock


\bibitem[Vasilisky et~al\mbox{.}(2023)]%
        {Vasilisky+al:23a}
\bibfield{author}{\bibinfo{person}{Ziv Vasilisky}, \bibinfo{person}{Oren Kurland}, \bibinfo{person}{Moshe Tennenholtz}, {and} \bibinfo{person}{Fiana Raiber}.} \bibinfo{year}{2023}\natexlab{}.
\newblock \showarticletitle{Content-Based Relevance Estimation in Retrieval Settings with Ranking-Incentivized Document Manipulations}. In \bibinfo{booktitle}{\emph{Proceedings of ICTIR}}. \bibinfo{pages}{205--214}.
\newblock


\bibitem[Wang et~al\mbox{.}(2024)]%
        {wang_text_2024}
\bibfield{author}{\bibinfo{person}{Liang Wang}, \bibinfo{person}{Nan Yang}, \bibinfo{person}{Xiaolong Huang}, \bibinfo{person}{Binxing Jiao}, \bibinfo{person}{Linjun Yang}, \bibinfo{person}{Daxin Jiang}, \bibinfo{person}{Rangan Majumder}, {and} \bibinfo{person}{Furu Wei}.} \bibinfo{year}{2024}\natexlab{}.
\newblock \bibinfo{title}{Text {Embeddings} by {Weakly}-{Supervised} {Contrastive} {Pre}-training}.
\newblock
\newblock
\urldef\tempurl%
\url{http://arxiv.org/abs/2212.03533}
\showURL{%
\tempurl}
\newblock
\shownote{arXiv:2212.03533 [cs]}.


\bibitem[Wang et~al\mbox{.}(2022)]%
        {Wang+al:22a}
\bibfield{author}{\bibinfo{person}{Yumeng Wang}, \bibinfo{person}{Lijun Lyu}, {and} \bibinfo{person}{Avishek Anand}.} \bibinfo{year}{2022}\natexlab{}.
\newblock \showarticletitle{{BERT} Rankers are Brittle: {A} Study using Adversarial Document Perturbations}. In \bibinfo{booktitle}{\emph{Proceedings of {ICTIR}}}. \bibinfo{pages}{115--120}.
\newblock


\bibitem[Wu et~al\mbox{.}(2022a)]%
        {Wu+al:22b}
\bibfield{author}{\bibinfo{person}{Chen Wu}, \bibinfo{person}{Ruqing Zhang}, \bibinfo{person}{Jiafeng Guo}, \bibinfo{person}{Wei Chen}, \bibinfo{person}{Yixing Fan}, \bibinfo{person}{Maarten de Rijke}, {and} \bibinfo{person}{Xueqi Cheng}.} \bibinfo{year}{2022}\natexlab{a}.
\newblock \showarticletitle{Certified Robustness to Word Substitution Ranking Attack for Neural Ranking Models}. In \bibinfo{booktitle}{\emph{Proceedings of CIKM}}. \bibinfo{pages}{2128--2137}.
\newblock


\bibitem[Wu et~al\mbox{.}(2022b)]%
        {Wu+al:22a}
\bibfield{author}{\bibinfo{person}{Chen Wu}, \bibinfo{person}{Ruqing Zhang}, \bibinfo{person}{Jiafeng Guo}, \bibinfo{person}{Maarten de Rijke}, \bibinfo{person}{Yixing Fan}, {and} \bibinfo{person}{Xueqi Cheng}.} \bibinfo{year}{2022}\natexlab{b}.
\newblock \bibinfo{title}{PRADA: Practical Black-Box Adversarial Attacks against Neural Ranking Models}.
\newblock
\newblock
\showeprint[arxiv]{2204.01321}


\bibitem[Wu et~al\mbox{.}(2024)]%
        {wu_result_2024}
\bibfield{author}{\bibinfo{person}{Haolun Wu}, \bibinfo{person}{Yansen Zhang}, \bibinfo{person}{Chen Ma}, \bibinfo{person}{Fuyuan Lyu}, \bibinfo{person}{Bowei He}, \bibinfo{person}{Bhaskar Mitra}, {and} \bibinfo{person}{Xue Liu}.} \bibinfo{year}{2024}\natexlab{}.
\newblock \bibinfo{title}{Result {Diversification} in {Search} and {Recommendation}: {A} {Survey}}.
\newblock
\newblock
\urldef\tempurl%
\url{http://arxiv.org/abs/2212.14464}
\showURL{%
\tempurl}
\newblock
\shownote{arXiv:2212.14464 [cs]}.


\bibitem[Zhai and Lafferty(2001)]%
        {Zhai+Lafferty:01a}
\bibfield{author}{\bibinfo{person}{Chengxiang Zhai} {and} \bibinfo{person}{John~D. Lafferty}.} \bibinfo{year}{2001}\natexlab{}.
\newblock \showarticletitle{A Study of Smoothing Methods for Language Models Applied to Ad Hoc Information Retrieval}. In \bibinfo{booktitle}{\emph{Proceedings of SIGIR}}. \bibinfo{pages}{334--342}.
\newblock


\end{thebibliography}

\end{document}